\documentclass[aps,english,superscriptaddress,preprintnumbers,reprint,footinbib,amsmath,amssymb,pra]{revtex4-2}
\usepackage{graphicx}
\usepackage[percent]{overpic}
\usepackage[version=3]{mhchem}
\usepackage{graphicx}
\usepackage{dcolumn}
\usepackage{bm}
\usepackage{hyperref}
\hypersetup{
    colorlinks=true,
    linkcolor=blue,
    filecolor=blue,
    urlcolor=blue,
   citecolor=blue,}
\usepackage{enumitem}
\usepackage{color}
\usepackage{amsfonts}
\usepackage{comment}

\begin{document}
\title{Quantum phase transitions and cat states in cavity-coupled quantum dots}

\author{Valerii K. Kozin}
\affiliation{Department of Physics, University of Basel, Klingelbergstrasse 82, CH-4056 Basel, Switzerland}
\author{Dmitry Miserev}
\affiliation{Department of Physics, University of Basel, Klingelbergstrasse 82, CH-4056 Basel, Switzerland}
\author{Daniel Loss}
\affiliation{Department of Physics, University of Basel, Klingelbergstrasse 82, CH-4056 Basel, Switzerland}
\author{Jelena Klinovaja}
\affiliation{Department of Physics, University of Basel, Klingelbergstrasse 82, CH-4056 Basel, Switzerland}

\date{\today}

\begin{abstract}
We study double quantum dots coupled to a quasistatic cavity mode with high mode-volume compression allowing for strong light-matter coupling.  Besides the cavity-mediated interaction, electrons in different double quantum dots interact with each other via dipole-dipole (Coulomb) interaction. There is a first-order cavity-induced ferroelectric quantum phase transition when the attractive dipolar interaction is smaller than the critical value defined by the energy splitting in DQDs and a smooth transition, otherwise. We show that, in the smooth transition region, both the ground and the first excited states of an array of double quantum dots are cat states.  Such states are actively discussed as high-fidelity qubits for quantum computing, and thus our proposal provides a platform for semiconductor implementation of such qubits. We also calculate gauge-invariant observables such as the net dipole moment, the optical conductivity, and the absorption spectrum beyond the semiclassical approximation. The results are robust against cavity losses and variations of system parameters.
\end{abstract}

\maketitle

\section{Introduction}
Placing condensed matter systems in an optical cavity is a promising way of engineering new correlated states of matter via the interaction with quantum fluctuations of the cavity field~\cite{cavQuantMatReview}. 
The main experimental challenge is to achieve the ultrastrong light-matter coupling regime~\cite{RevModPhys.91.025005,FriskKockum2019} that can be reached by external driving~\cite{FloquetReview,PhysRevLett.116.176401,PhysRevB.81.165433,PhysRevB.79.081406,Lindner2011,Dehghani2015,doi:10.1126/science.1239834,kozin_topo_2D_QR,PhysRevB.97.155434}, by tuning the cavity in a plasmon or an exciton-polariton resonance~\cite{MicrocavitiesKavokin,kozin_2021_pol_ring_exp,PhysRevB.98.125115,PhysRevA.100.043610}, or by compressing  the mode volume in specially designed resonators~\cite{ModeCompressionPRB,ModeCompressionNanoLett} 
that can be viewed as $LC$-circuits~\cite{PhysRevX.4.041031,PhysRevB.85.045304} with a single discrete quasistatic mode, whose frequency $\omega_0 = 1 / \sqrt{LC}$ is not constrained by the resonator dimensions. 
When the light-matter coupling is strong enough, then even in the ground state the vacuum fluctuations can radically modify electron systems~\cite{FaistHallBreakdown,PhysRevB.99.235156,PhysRevX.10.041027,kozin2024cav_SC}. This phenomenon fosters a qualitatively new class of condensed-matter platforms with strongly correlated light-matter excitations.

Superradiance, initially described by R. H. Dicke~\cite{PhysRev.93.99,Kirton2018}, has garnered significant attention to coupled light-matter systems ever since.
There exist various effective models describing cavity-coupled electron systems, known as extended and generalized Dicke models, see, e.g.,  Refs.~\cite{PhysRevA.8.1440,PhysRevA.97.043820,PoliniGraphene,kurlov2023generalization}. 
An important restriction to such effective models is gauge invariance that must be preserved~\cite{Nataf2010,PhysRevB.102.125137,10.21468/SciPostPhys.15.3.113}.
Originally, the main signature of the superradiant Dicke phase transition was a photon condensate, the macroscopic occupation of the cavity mode that is not gauge-invariant~\cite{RevModPhys.94.045003}. 
Nevertheless, the quantum phase transition (QPT) is present and equivalent to the ferroelectric phase transition (FPT), resulting in ordered electric dipole moments, see Refs.~\cite{PhysRevLett.125.143603,PhysRevLett.112.073601}.
Important, the FPT is only possible if the Coulomb interaction between the dipoles is included~\cite{Keeling2007,PhysRevLett.35.432,PhysRevA.97.043820,PhysRevLett.125.143603,10.21468/SciPostPhys.9.5.066}.

\begin{figure}[b]
\includegraphics[width=1\columnwidth]{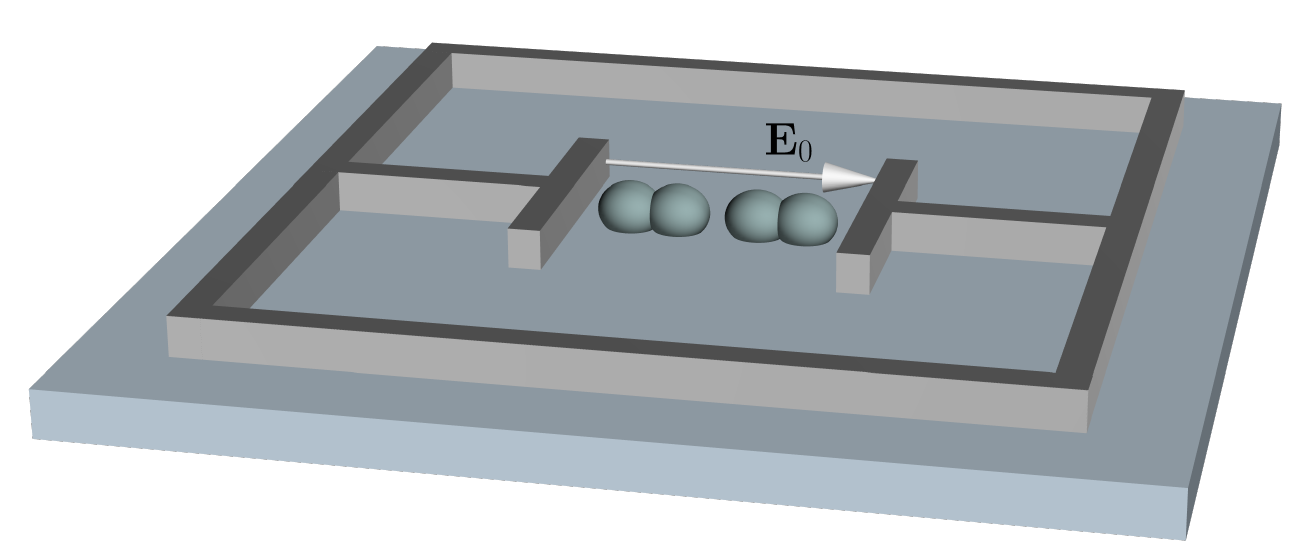}
\put(-245,94){(a)}\\
\includegraphics[width=0.88\columnwidth]{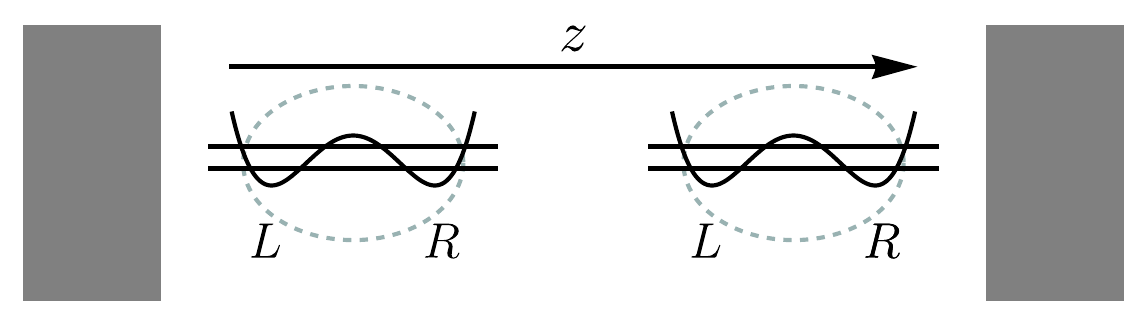}
\put(-231,48){(b)}
\caption{\label{fig:1} 
(a) Sketch of the system: 
Two DQDs embedded in a split-ring resonator. The cavity mode is polarized along the arrow, $\mathbf{E}_0=(0,0,E_0)$. (b) Two double-well potentials for each DQD. The two lowest energy levels are shown. The two minima of each DQD are marked with L (Left) and R (Right), respectively. The DQD axes are aligned along the $z$ axis. }
\end{figure}

In this work, we consider a few cavity-coupled double quantum dots (DQDs) with a Coulomb interaction between them, the only non-trivial part of which is the electric dipole-dipole interaction.
{Choosing a geometry where the dipolar interaction between DQDs is attractive, we find either a first order QPT or a smooth transition depending on the relative strength of the Coulomb and light-matter interactions}, leading to ordered phases of the electric dipole moments.
The ground and the first excited states are cat states in the smooth transition region.
In particular, this is true already for two cavity-coupled DQDs with attractive dipole-dipole (Coulomb) interaction.
We suggest such systems as possible semiconductor candidates for a self-correcting cat qubit~\cite{Xu2023,Gertler2021} and a realistic platform to study cavity-induced QPTs. Here we calculate the net dipole moment, the optical conductivity, and the absorption spectrum all of which are gauge-invariant.

\section{Theoretical Model} 
A few identical singly-occupied DQDs are oriented along the line connecting the capacitor plates as shown in Fig.~\ref{fig:1}. 
Due to the Coulomb repulsion, DQDs interact with each other directly via the electric dipole-dipole interaction. 
The double-well shape of the confining potential of each DQD allows us to truncate electron energy levels by the lowest two as long as the higher states are far detuned~\cite{PhysRevA.98.053819,PhysRevB.101.205140}. Such an electronic system is described by the following Hamiltonian, 
\begin{eqnarray}
 {H}_{\mathrm{el}} & = & - \frac{\Delta}{2} \sum_{i=1}^N \left({c}_{i,L}^\dagger{c}_{i,R}+\text{h.c.}\right) \nonumber \\
 & + & \sum_{i>j}^N \frac{U_{ij}}{2} {d}_{i,z}{d}_{j,z} - \frac{V_b}{2} \sum_{i=1}^N {d}_{i, z} \, , 
\end{eqnarray}
where ${c}^\dagger_{i,L/R}$ (${c}_{i,L/R}$) are the electron creation (annihilation) operators for the two sites ($L/R$) of the $i^{\rm th}$ DQD, 
$N$ is the number of DQDs,
$V_b$ is the bias in each DQD, $\Delta$ is the DQD level hybridization,
and ${d}_{i,z}={c}^\dagger_{i,R}{c}_{i,R}-{c}^\dagger_{i,L}{c}_{i,L}$ is the electric dipole operator. Spin indices are suppressed. 
The Coulomb interaction is reduced to the dipole-dipole interaction here due to the two-level truncation of each singly-occupied DQD.
The dipolar interaction strength between two DQDs is derived in Appendix~\ref{dipole_int_deriv}
\begin{eqnarray}
	&& U_{ij} \approx \frac{e^2 b^2}{2 \varepsilon} \frac{|\mathbf{r}_{ij}|^2-3(\mathbf{r}_{ij}\cdot\mathbf{e}_z)^2}{|\mathbf{r}_{ij}|^5} \, ,
\end{eqnarray}
where $\bm{r}_{ij} = r_{ij} \mathbf{e}_z$ is the distance vector between two DQD centers, 
$\varepsilon$  the dielectric constant,
$e < 0$ the elementary charge,
$b$ the DQD length, and $e b/2$ the dipole matrix element between the lowest two levels of a DQD. 
If DQDs are assembled along the capacitor axis $z$, the dipole-dipole interaction is attractive, $U_{ij} = -(e b)^2/(\varepsilon r_{ij}^3) < 0$.
Screening of $U_{ij}$ due to proximity to the capacitor plates does not affect the sign of $U_{ij}$ but only slightly modifies its absolute value. 
In what follows, we mostly focus on two ($N = 2$) DQDs.
In this case, only the $U_{12} \equiv U$ matrix element of the dipole-dipole interaction is important.
Throughout the paper, we use cgs-units and also set the Planck and Boltzmann constants to unity, $\hbar = k_B = 1$.

All DQDs are coupled to a single quantized quasistatic $LC$-cavity mode~\cite{PhysRevX.8.041018, PhysRevX.12.031004, Valmorra2021, PhysRevB.88.125312,PhysRevLett.130.233602,epxerimentGateDefinedQDandSplitRing}. 
The electric field of the cavity mode is almost completely localized in the capacitor and polarized along the DQDs, see Fig.~\ref{fig:1}. 
The corresponding vector-potential operator ${A}_z$ is given by
\begin{eqnarray}
	&& {A}_z = i \frac{E_0}{\omega_0} ({a}^\dagger-{a}) \equiv i\sqrt{\frac{2\pi}{\varepsilon V_\mathrm{eff}\,\omega_0}}({a}^\dagger-{a}) \, ,
\end{eqnarray}
where $a$ ($a^\dagger$) is the annihilation (creation) operator of the cavity mode
with frequency $\omega_0$, $E_0$  the amplitude of the electric field fluctuations, and $V_\text{eff}$ the effective mode volume.

\begin{figure}[t]
\begin{overpic}[scale=0.45]{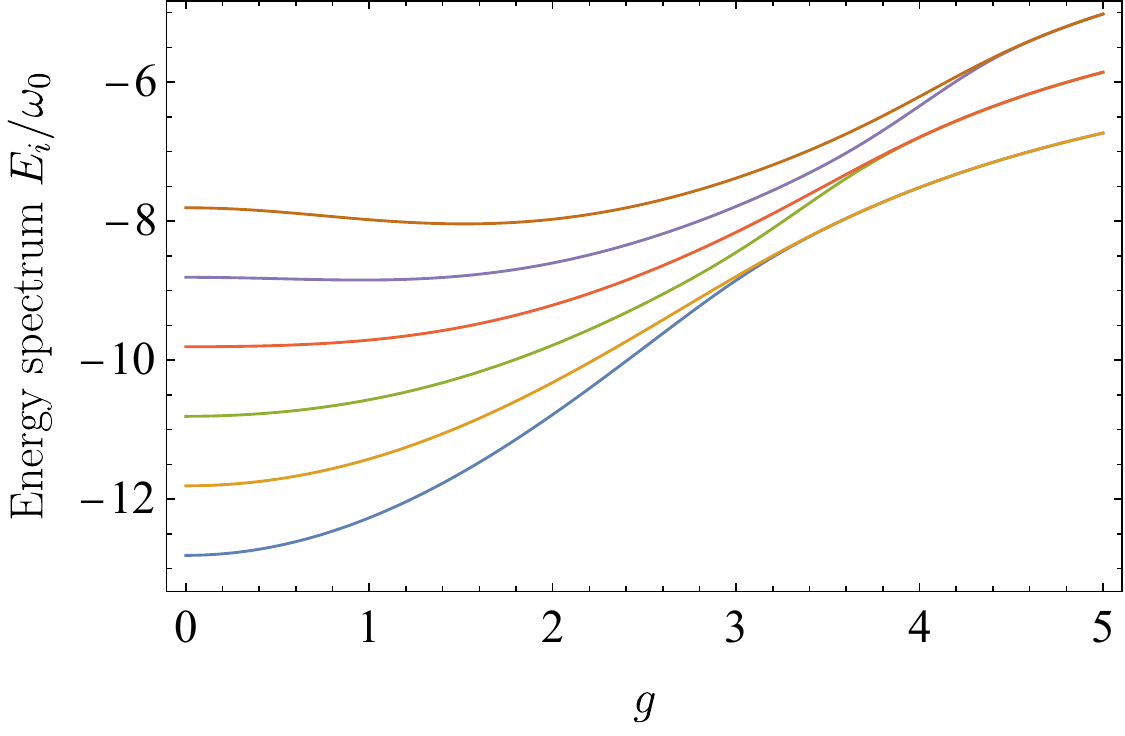}
     \put(57,12){\includegraphics[scale=0.133]{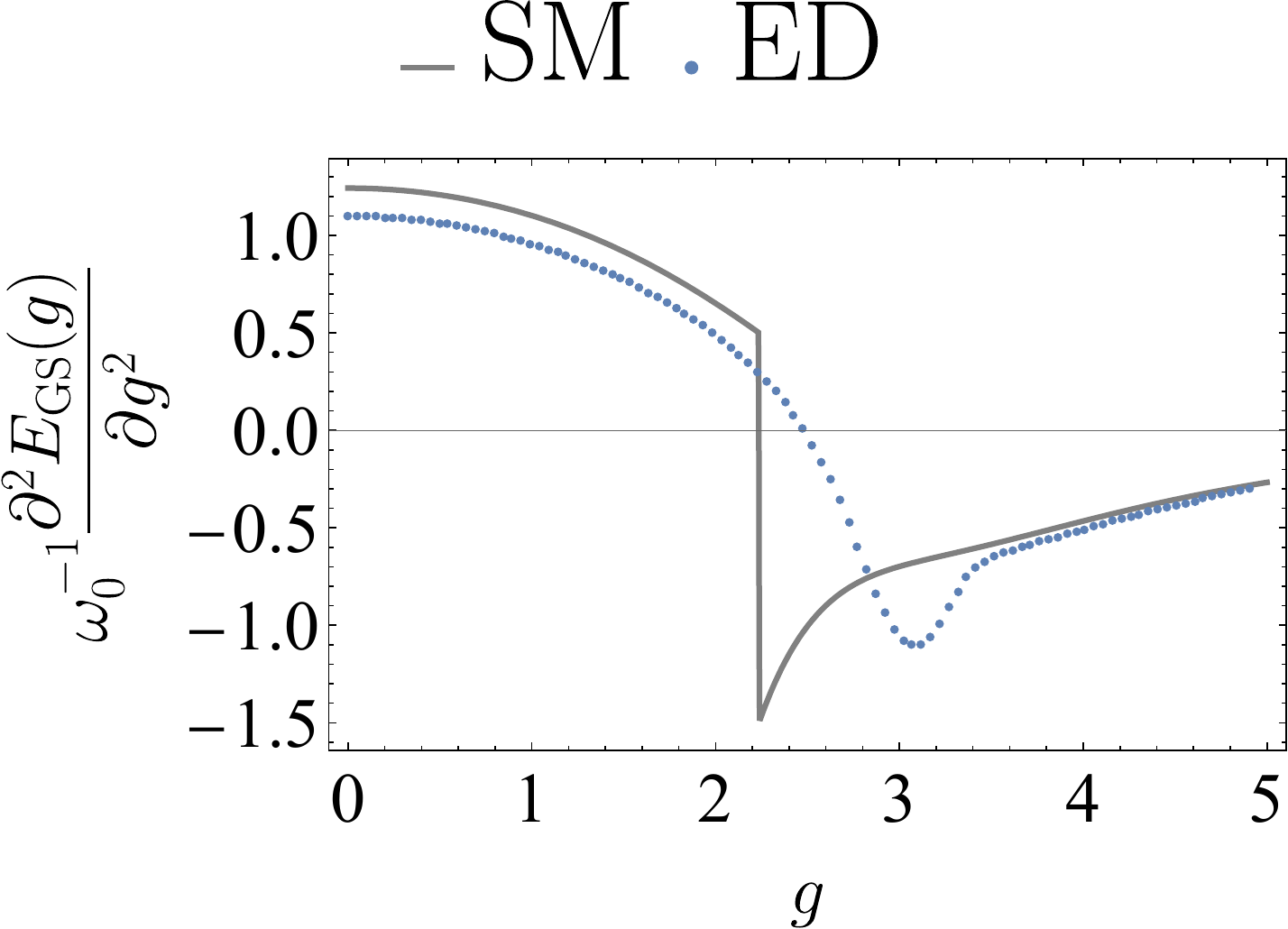}}
\end{overpic}
\put(-243,150){(a)}\\
\begin{overpic}[scale=0.45]{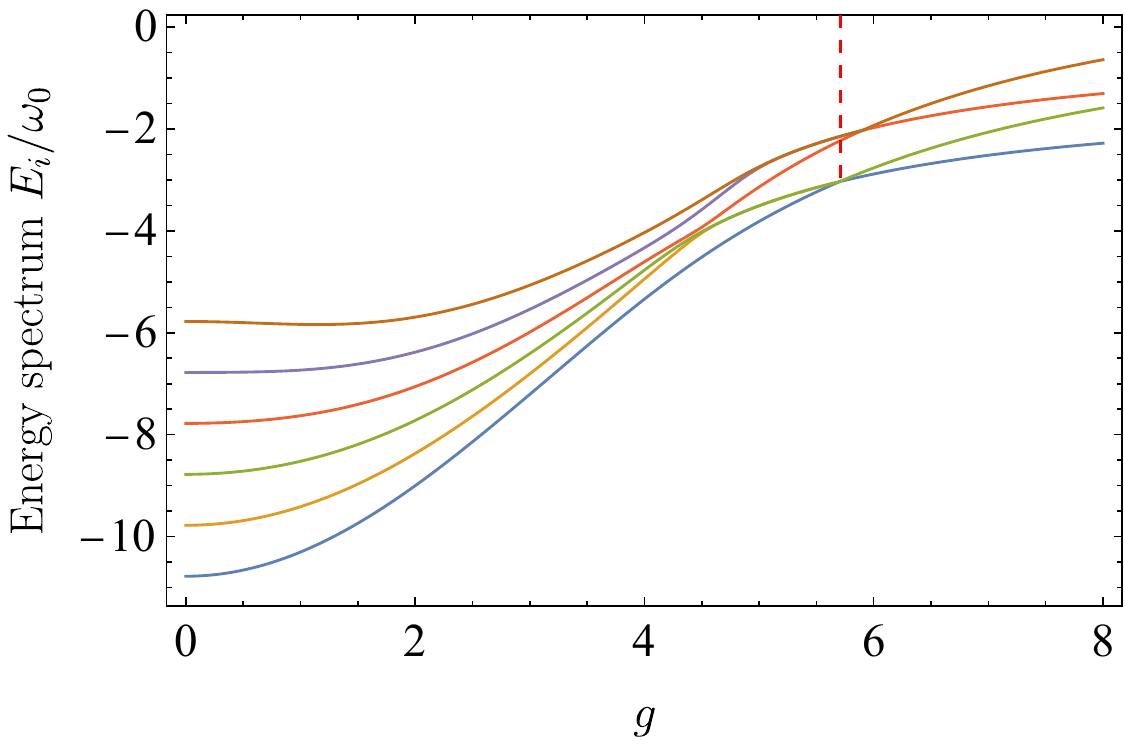}
     \put(54,12.5){\includegraphics[scale=0.18]{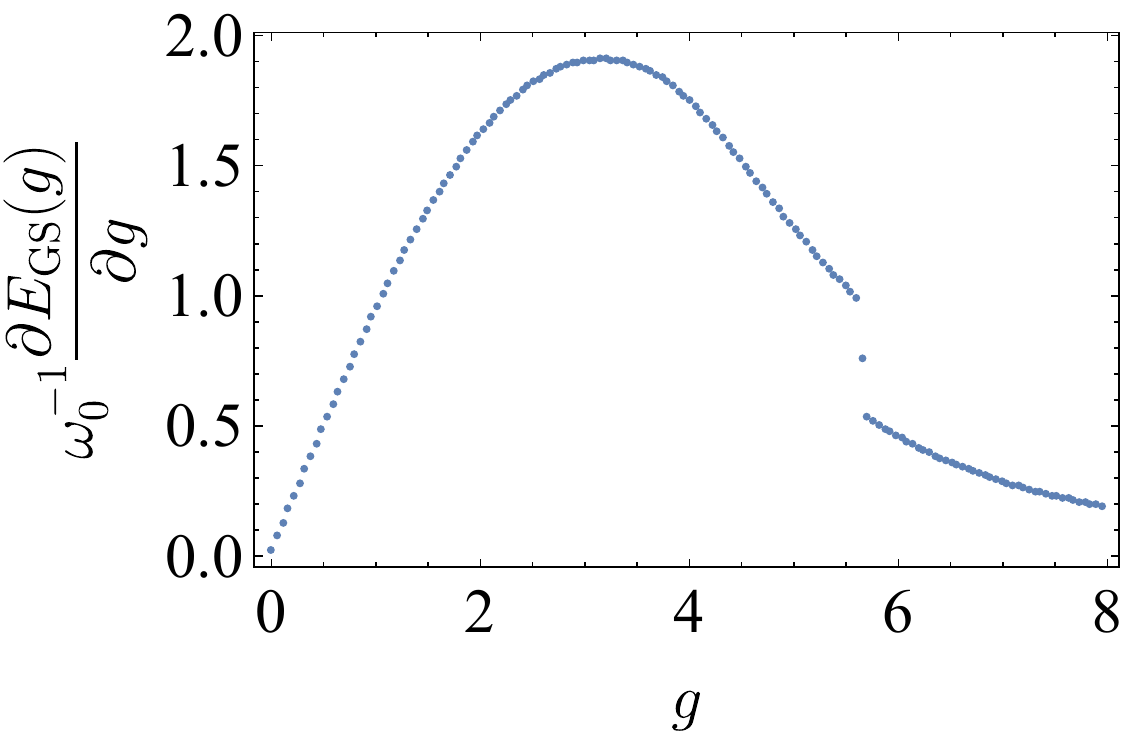}}
     \put(16,42){\includegraphics[scale=0.126]{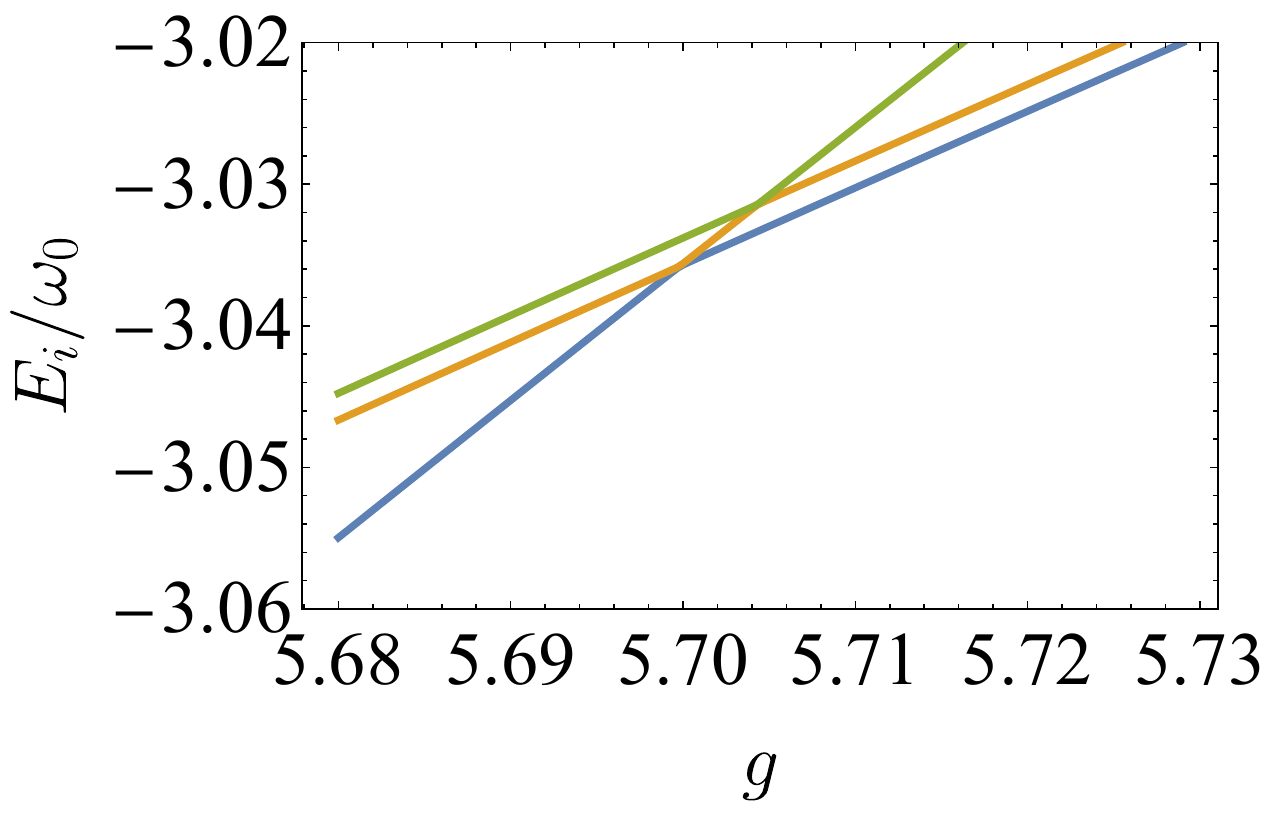}}
\end{overpic}
\put(-243,150){(b)}
\caption{\label{fig:energy_spectum} 
{Numerically exact six lowest energy levels of $H$, Eq.~(\ref{eq:Peierls_Hamilt}), are shown for $\omega_0/ \Delta = 0.1$ and attractive dipolar interaction
(a) $U/\omega_0=-5$, (b) $U/\omega_0=-1.5$ as a function of light-matter coupling $g$. 
Panel (a): energy levels merge pairwise, indicating the smooth transition to the ordered state. The inset shows the second derivative of the ground state energy with respect to the light-matter coupling $g$, where the dotted line is obtained by exact diagonalization (ED), the solid line corresponds to the semiclassical (SM) approximation.  Panel (b) shows the level crossing corresponding to a first-order ferroelectric QPT, indicated by the vertical dashed red line. The left inset in (b) shows the zoomed-in level crossing region (where we introduced a small ratio $V_b/\omega_0=0.5\cdot10^{-3}$ to identify the two otherwise degenerate levels). The right inset in (b) shows the first derivative of the ground state energy $\partial E_{\text{GS}}(g)/\partial g$ obtained by ED that is discontinuous at the QPT. }}
\end{figure}

We {describe the coupling of the DQDs to the cavity via the Peierls substitution}, 
\begin{align}
 & {H}_0 = \omega_0 {a}^\dagger {a} - V_b {S}_z + U {S}_z^2 , \label{H0} \\
 &{H} = {H}_0 - \frac{\Delta}{2} \left(e^{g (a - a^\dagger)} {S}_+ + e^{-g (a - a^\dagger)} {S}_- \right)  ,\label{eq:Peierls_Hamilt}\\
 & g = \sqrt{\frac{W} {\omega_0}} \, , \hspace{5pt} W =   \frac{2 \pi e^2 b^2}{\varepsilon V_\text{eff}}, \label{g} 
\end{align}
where ${S}_\beta = 1/2\sum_{i = 1}^N \sigma_{i,\beta}$ is the orbital pseudospin of the system, $\sigma_{i, \beta}$ is the Pauli matrix corresponding to the $i^{\rm th}$ DQD, $\beta \in \{x, y, z\}$,
$g$ is the dimensionless light-matter coupling constant,
$U_{ij} = U$. 
The  operators ${S}_z$ and ${S}_\pm = {S}_x \pm i {S}_y$ satisfy the standard spin algebra: $[{S}_\pm, {S}_z] = \mp {S}_\pm$, $[{S}_+, {S}_-] = 2 {S}_z$.
The pseudospin ${\bm{S}}$ describes the collective orbital degree of freedom in the DQD array. 
For example, the total dipole moment operator maps onto ${S}_z$: $\sum_{i=1}^N{d}_{i,z}\to 2 {S}_z$.

\begin{figure}[t]
\includegraphics[width=0.76\columnwidth]{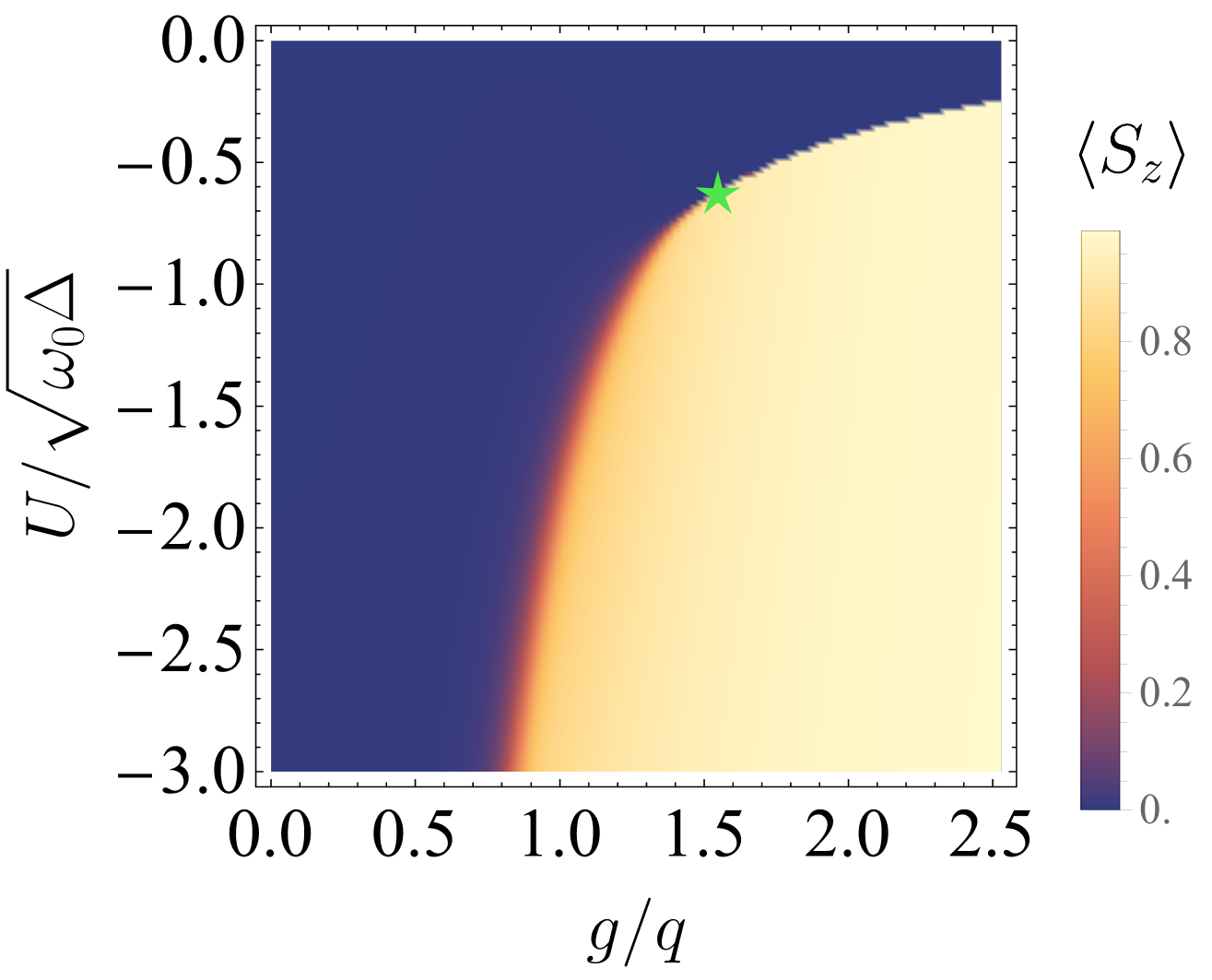}
\put(-187,140){(a)} \\
\includegraphics[width=0.76\columnwidth]{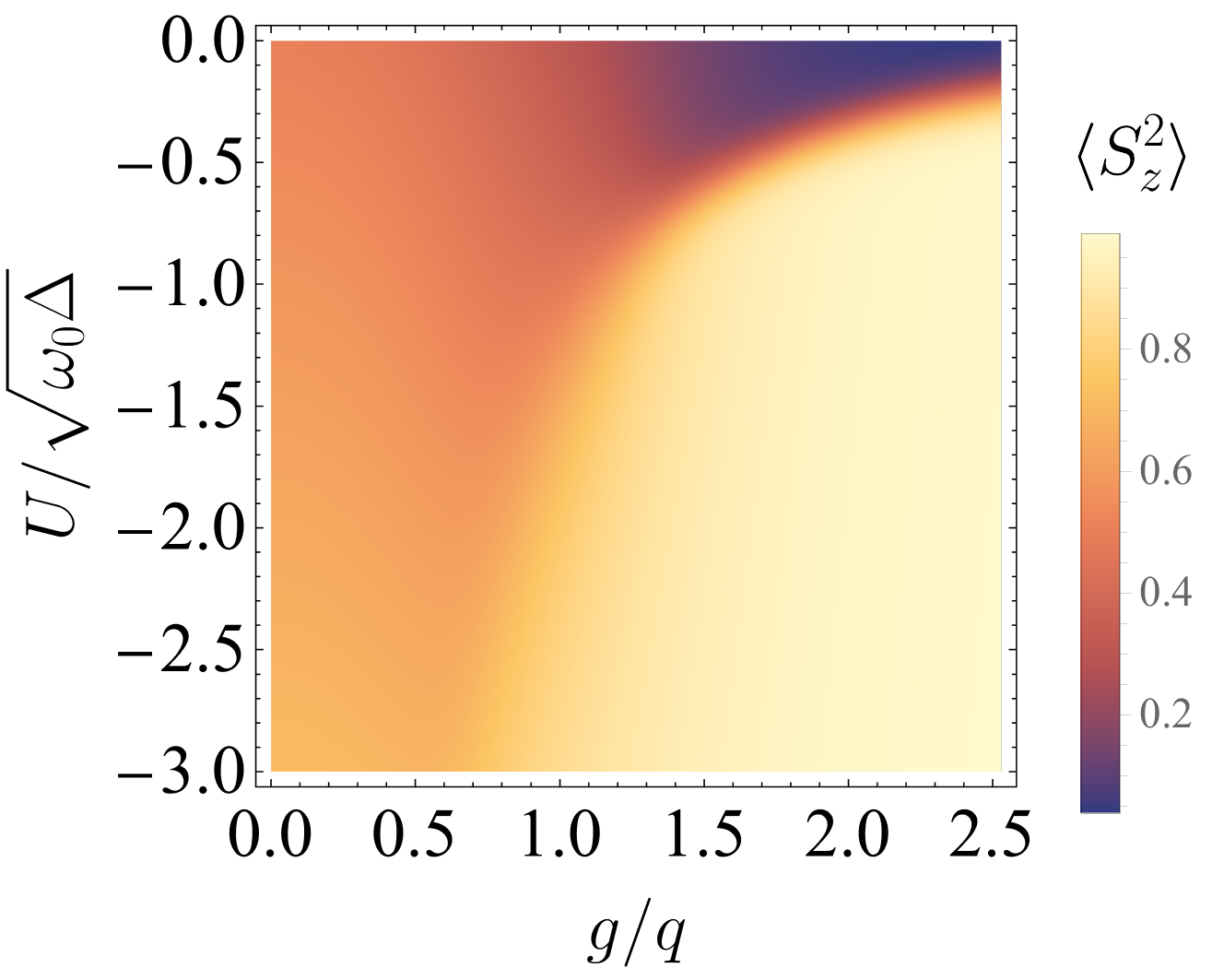}
\put(-187,140){(b)}
\caption{\label{fig:Sz_map} 
(a) Phase diagram of the numerically exact net dipole moment $\langle{S}_z\rangle$ for $N = 2$ DQDs at $\omega_0/\Delta =1/q^2= 0.1$ and zero temperature as a function of the dipole interaction strength and light-matter coupling constant. An infinitesimal symmetry-breaking field $-V_b {S}_z$ with $V_b/\sqrt{\omega_0\Delta}=10^{-3}$ is introduced. 
(b) Phase diagram of $\langle{S}_z^2\rangle$ at finite temperature $T/\omega_0=0.1$.}
\end{figure}

First, we diagonalize the Hamiltonian $H$, see Eq.~\eqref{eq:Peierls_Hamilt}, numerically for $N = 2$ cavity-coupled DQDs at zero bias $V_b = 0$, {truncating the photon Hilbert space at 200 photons to ensure convergence. 
First six energy levels are shown in Fig.~\ref{fig:energy_spectum}. The spectrum demonstrates either continuous coalescence of the energy levels corresponding to a smooth transition to the ferroelectric phase, Fig.~\ref{fig:energy_spectum}(a), or a level crossing indicating the first-order ferroelectric QPT, Fig.~\ref{fig:energy_spectum}(b).} 

The zero-temperature phase diagram represented by a 2D plot of the net dipole moment $\langle{S}_z\rangle$ as a function of the dipole interaction strength and light-matter coupling constant at $\omega_0/\Delta = 0.1$ [see Fig.~\ref{fig:Sz_map}(a)]
shows the first-order QPT at $|U/\sqrt{\omega_0\Delta}|<U_c/\sqrt{\omega_0\Delta}\approx 0.2 \sqrt{\Delta/\omega_0}$, $g>g_c\approx1.5\sqrt{\Delta/\omega_0}$ and the smooth transition otherwise, separated by the critical point ($g_c$,$U_c$) marked by the star. 
It's position remains fixed if plotted in the coordinates $(g/q,U/\sqrt{\omega_0\Delta})$, where $q=\sqrt{\Delta/\omega_0}$, in agreement with the mean-field analysis shown in Appendix~\ref{semiclass_an}.
At finite temperature, the QPT turns into a smooth transition, see the density plot of $\langle{S}_z^2\rangle$ in Fig.~\ref{fig:Sz_map}(b). 
Here, we used that $\langle S_z^2\rangle$ does not require a small symmetry breaking field $V_b$ which is useful for the finite-temperature analysis.
We point out that $\langle S_z^2\rangle$ is meaningful for $N \ge 2$ DQDs.
We also stress that there is no QPT at $g=0$ implying that this is a cavity-induced phenomenon.

\section{Semiclassical decoupling}

In order to gain physical insight into our numerical results, {we analyze the system in the quasi-thermodynamic limit $\omega_0 \ll \Delta$ (the limit of the classical oscillator), see Refs.~\cite{PhysRevA.85.043821,PhysRevA.87.013826,PhysRevLett.115.180404,puebla2017probing} for details. Our results remain qualitatively the same even when $\Delta\sim\omega_0$, see Appendix~\ref{non_equiv_QD}.}
The photonic semiclassical decoupling is reminiscent of the length-gauge formulation of the problem~\cite{Power1957,Woolley1971,Dmytruk2021,PhysRevB.108.085410},
\begin{align}
&\hspace{-3pt} {H}_{\text{D}} = {\mathcal{U}} {H} {\mathcal{U}}^\dagger = {H}_{\text{sm}} + \delta {V} , \label{HD} \\
&\hspace{-3pt} {H}_{\text{sm}} = \omega_0 {a}^\dagger {a} -V_b{S}_z-\Delta{S}_x + U{S}_z^2 + g^2 \omega_0 \left( \delta {S}_z\right)^2 , \label{meanfield} \\
&\hspace{-3pt} \delta {V} = - g \omega_0 \, \delta {S}_z \left( {a} + {a}^\dagger\right) , \label{Vmf}
\end{align}
where ${H}$ is given by Eq.~(\ref{eq:Peierls_Hamilt}),
${\mathcal{U}}=\exp{\left[g \, \delta {S}_z({a}^\dagger-{a})\right]}$, $\delta {S}_z = {S}_z - \langle {S}_z \rangle$, $\langle {S}_z \rangle$ is the average of orbital pseudospin ${S}_z$ over the ground state of the semiclassical Hamiltonian ${H}_{\text{sm}}$. 
The perturbation $\delta {V}$ accounts for quantum corrections beyond the semiclassical approximation.
In contrast to conventional mean-field treatment where both, the photons and the pseudospin, are treated as classical objects, see e.g. Ref.~\cite{PhysRevA.85.043821},
the orbital pseudospin ${\bm{S}}$ in our work remains quantum because we apply our results to a small number of DQDs.

The Hamiltonian ${H}_D$ commutes with ${\bm{S}}^2$, so we consider states with definite orbital pseudospin $S$. 
Single DQD corresponds to $N = 1$ and $S = 1/2$, this case is known as the quantum Rabi model~\cite{PhysRevLett.115.180404,PhysRevLett.131.113601}.
In case of $N = 2$ DQDs, see Fig.~\ref{fig:1}, $S$ can be either $0$ or $1$.
The $S = 0$ state does not couple to the antenna.
If $S = 1$, the semiclassical Hamiltonian [Eq.~(\ref{meanfield})] can be diagonalized analytically, see Appendix~\ref{semiclass_an}.
Here, we show the semiclassical ground-state energy, $E_{\text{sm}}$, of 
two cavity-coupled DQDs at $V_b = 0$ (symmetric DQDs):
\begin{align}
&E_{\text{sm}} = - \frac{2}{3} \sqrt{P} \cos \left[\arccos\left(Q/P^{3/2}\right)/3\right]\nonumber\\
& + \frac{2}{3}\left(U + g^2 \omega_0\right) + \omega_0 \alpha^2 , \label{Emf}
\end{align}
where $P$ and $Q$ are defined as follows, 
\begin{align}
&\hspace{-5pt} P = \left(U+g^2\omega_0\right)^2 +3\left(\Delta^2+4g^2\alpha^2\omega_0^2\right),\\
&\hspace{-5pt} Q = \left(U+g^2 \omega_0\right)\left[ \left(U + g^2 \omega_0 \right)^2 - 36 \alpha^2 g^2 \omega_0^2 + 9 \Delta^2/2\right] . \nonumber
\end{align}
Here, we introduced the parameter $\alpha = g \langle {S}_z \rangle$.
If $\alpha = 0$, the phase is trivial.
If $\alpha \ne 0$, the ground state is ferroelectric, i.e., it has a net dipole moment $\langle {S}_z \rangle \ne 0$.
The ferroelectric QPT  is first-order and $\langle {S}_z \rangle$ has a finite jump at the transition, see Fig.~\ref{fig:Sz_map}(a), whereas the second-order QPT predicted by the mean field turns into a smooth transition in the ED due the tunneling effect.

As $E_{\text{sm}}$ is an even function of $\alpha$ at $V_b = 0$ [Eq.~(\ref{Emf})], the ground state of ${H}_{\text{sm}}$ is two-fold degenerate at $\alpha \ne 0$.
This degeneracy is best seen from the symmetry $\mathcal{P} = \exp(i \pi {a}^\dagger {a} + i \pi {S}_x)$ of the transformed Hamiltonian ${H}_T = {T}(\alpha) {H}_D {T^\dagger}(\alpha)$ at $V_b = 0$, where ${T}(\alpha) = e^{\alpha ({a}^\dagger - {a})}$.
Note that ${T}(\alpha)$ is the optical displacement operator that creates the coherent state $|\alpha \rangle = {T}(\alpha) |0\rangle$, where $|0\rangle$ is the photonic vacuum of ${H}_\text{sm}$, see Eq.~(\ref{meanfield}).
While the symmetry breaking in this problem occurs only in the limit $\omega_0/\Delta \to 0$, we expect very small lifting of the degeneracy at any finite $\omega_0/\Delta \ll 1$ such that the parity symmetry $\mathcal{P}$ of the Hamiltonian ${H}_T$ is restored.
In particular, the ground state, $|\Psi_\text{G}\rangle$, and the first excited state, $|\Psi_\text{E1}\rangle$, of the Hamiltonian ${H}_T$ in the semiclassical approximation correspond to $\mathcal{P} = +1$ and $\mathcal{P} = -1$, respectively,
 \begin{align}
    & |\Psi_\text{G}\rangle = \mathcal{N}\left[\chi(\alpha) |\alpha\rangle  + \chi(-\alpha) |-\alpha\rangle\right],  \label{eq:cat_states_gs}\\
    & |\Psi_\text{E1}\rangle = \mathcal{N}\left[\chi(\alpha) |\alpha\rangle  - \chi(-\alpha) |-\alpha\rangle\right] ,
    \label{eq:cat_states_ex}
 \end{align}
 where $\alpha > 0$ corresponds to positive dipole moment $\langle {S}_z \rangle > 0$, $\chi(\pm \alpha)$ are the two lowest-energy eigenstates of ${H}_\text{sm}$, and $\mathcal{N}$ is the normalization factor.
 Indeed, we observe a finite splitting in the ferroelectric phase, see Fig.~\ref{fig:energy_spectum}(a).
Restoration of the parity symmetry $\mathcal{P}$ is due to the tunneling (instantons) between two semiclassical ground states~\cite{Vanshten1982}.

\begin{figure}[t]
  \begin{overpic}[scale=0.45]{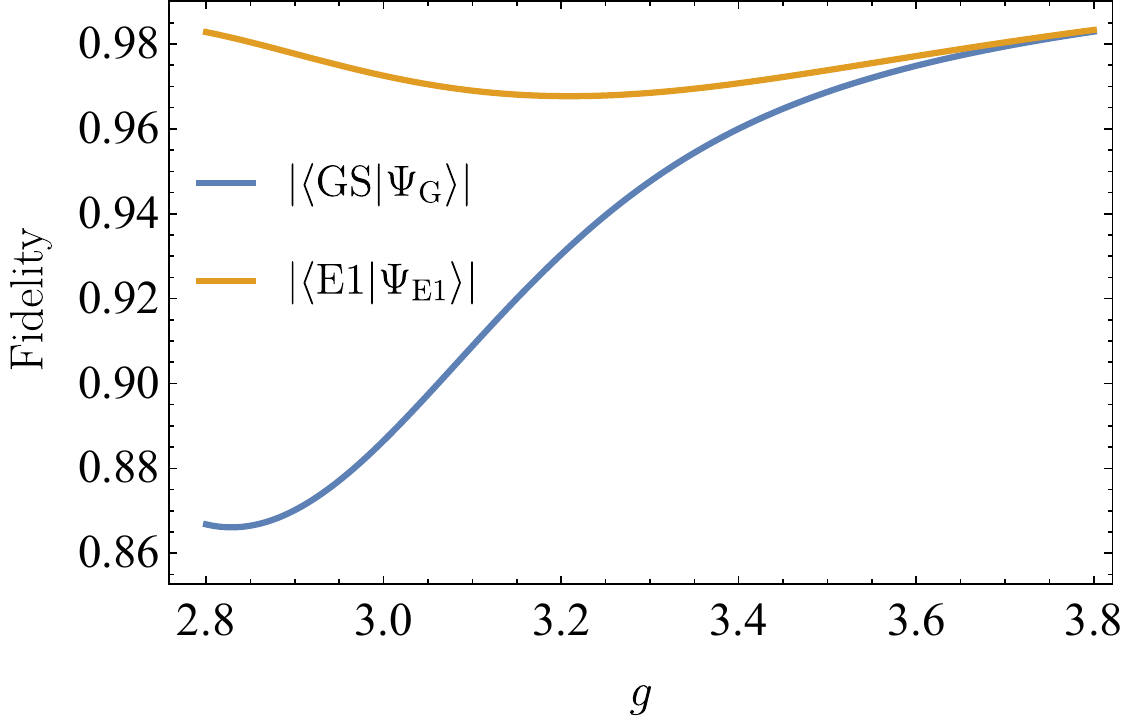}
     \put(51,13){\includegraphics[scale=0.19]{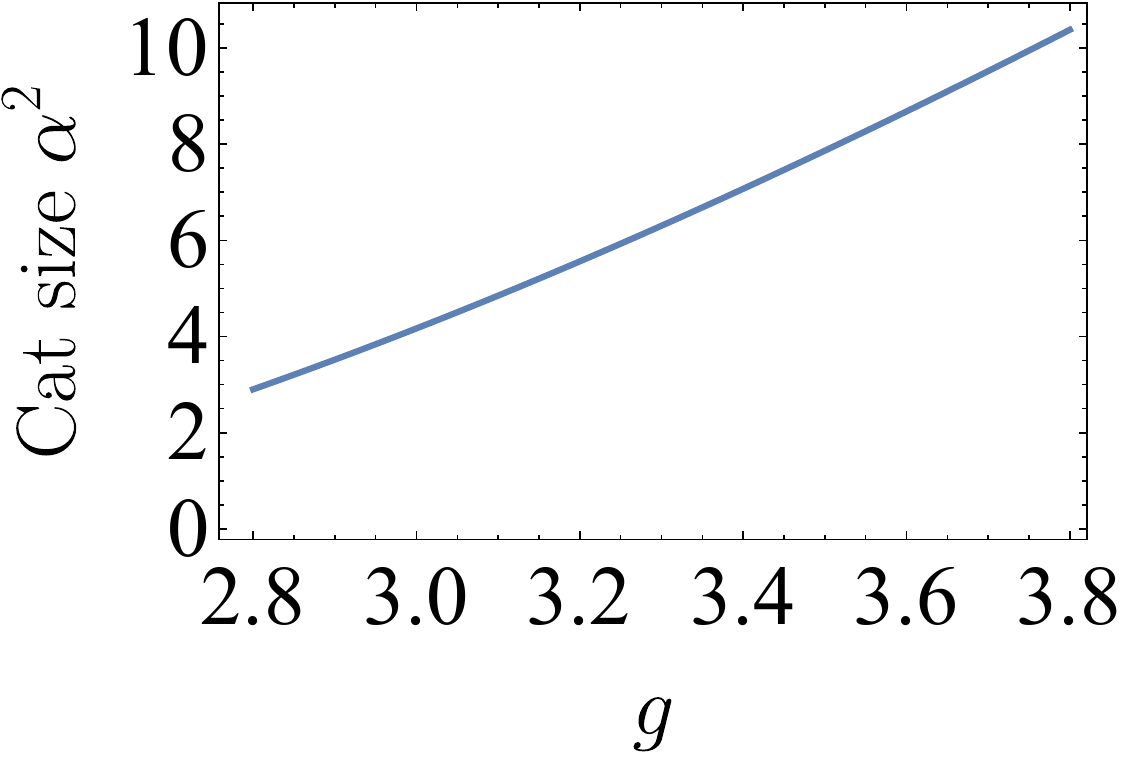}}
  \end{overpic}
\caption{\label{fig:fidelity} 
Fidelities $|\langle\text{GS}\lvert\Psi_{\text{G}}\rangle|$ 
and $|\langle\text{E1}\lvert\Psi_{\text{E1}}\rangle|$ in the smooth transition region as a function of light-matter coupling $g$, where $|\text{GS} \rangle$ and $|\text{E1}\rangle$ are the exact ground and the first excited states (in the length gauge), $\lvert\Psi_{\text{G}}\rangle$ and $\lvert\Psi_{\text{E1}}\rangle$ are the semiclassical cat states, see Eqs.~\eqref{eq:cat_states_gs} and  \eqref{eq:cat_states_ex}.
The semiclassical approximation is valid if the fidelities are close to one. 
The following parameters are used: the dipole-dipole interaction strength is $U/\omega_0 = -5$ and $\omega_0/\Delta=0.1$. 
The inset shows the ``cat size" $\alpha^2$.}
\end{figure}

The fidelities $|\langle\text{GS}\lvert\Psi_{\text{G}}\rangle|$ and $|\langle\text{E1}\lvert\Psi_{\text{E1}}\rangle|$  plotted in Fig.~\ref{fig:fidelity} as function of $g$ justify the semiclassical treatment in the ferroelectric phase,
where $|\text{GS} \rangle$ and $|\text{E1}\rangle$ are exact (numerical) ground and first excited states, $\lvert\Psi_{\text{G}}\rangle$ and $\lvert\Psi_{\text{E1}}\rangle$ are corresponding semiclassical cat states, see Eqs.~\eqref{eq:cat_states_gs} and \eqref{eq:cat_states_ex}.
This confirms the semiclassical result that the ground and the first excited states are two-component cat states. The parameter $\alpha^2$, being an increasing function of $g$ (see the inset in Fig.~\ref{fig:fidelity}), plays the role of the ``cat size''.  
The comparison between the semiclassically calculated phase diagram for the order parameter $\langle \hat{S}_z^2\rangle$ and the ED is shown in Appendix~\ref{semiclass_phase_diag}.
Two lowest energy levels become degenerate in the strong coupling limit $g \to +\infty$, see Fig.~\ref{fig:energy_spectum}, when the Schr\"{o}dinger cats become truly classical.
In order to use such a system as a cat qubit, a finite energy splitting is required {which corresponds to the smooth transition region and restricts the cat size $\alpha^2$}.
On the bright side, the cat states appearing at strong coupling ({recently observed in circuit QED~\cite{PhysRevLett.131.113601,Reglade2024}}) are robust to decoherence and can be harnessed to implement quantum gates with high fidelity~\cite{PhysRevLett.107.190402,wang2016holonomic}. 
We propose the cavity-coupled DQDs as a new solid-state platform for cat qubits {(without driving~\cite{PhysRevResearch.3.023088})}, as promising candidates for quantum computing~\cite{PRXQuantum.3.010329,PRXQuantum.4.020337,PhysRevA.106.022431}. 
In contrast to atomic systems (e.g., see~\cite{PhysRevLett.125.263606,PhysRevX.8.021027,Cai2021}), solid-state platforms are scalable and require much less stringent experimental conditions. {As shown in Appendix~\ref{non_equiv_QD}, the results are resilient to variations of the DQD parameters. Also, when the cavity losses are included within the Lindblad formalism, the phase transition is shown to  remain first-order as shown in Appendix~\ref{app_Lindblad}. The behaviour of the cat states is analyzed within the quantum jump (Monte Carlo) method revealing switching between the two cat states which gives rise to a finite coherence time of the cat qubit (see Appendix~\ref{app_MC}).}

\section{Optical conductivity and absorption spectrum} 

\begin{figure}[h]
\hspace*{0.5cm}\includegraphics[width=0.76\columnwidth]{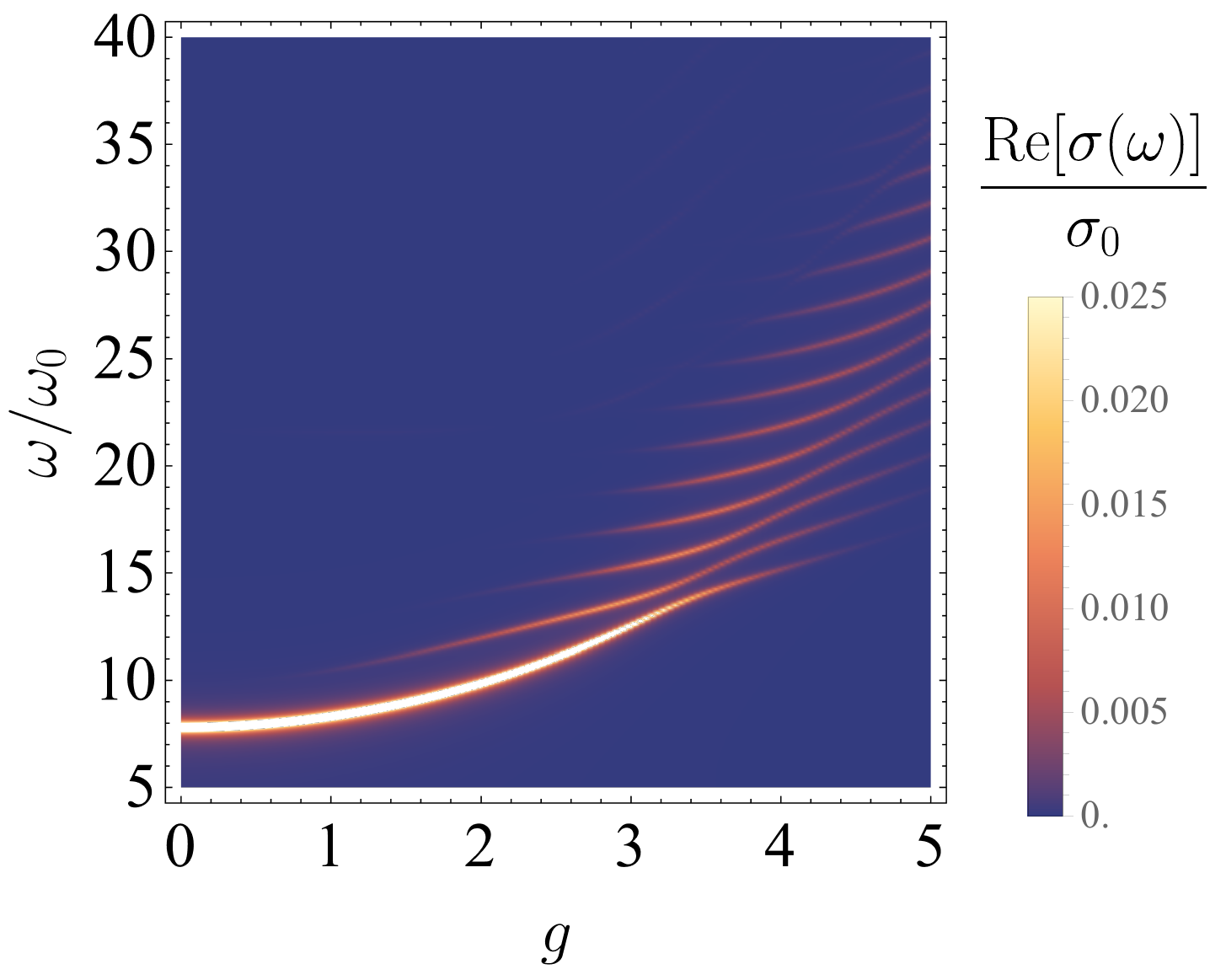}
\put(-189,140){(a)}\\
\includegraphics[width=0.71 \columnwidth]{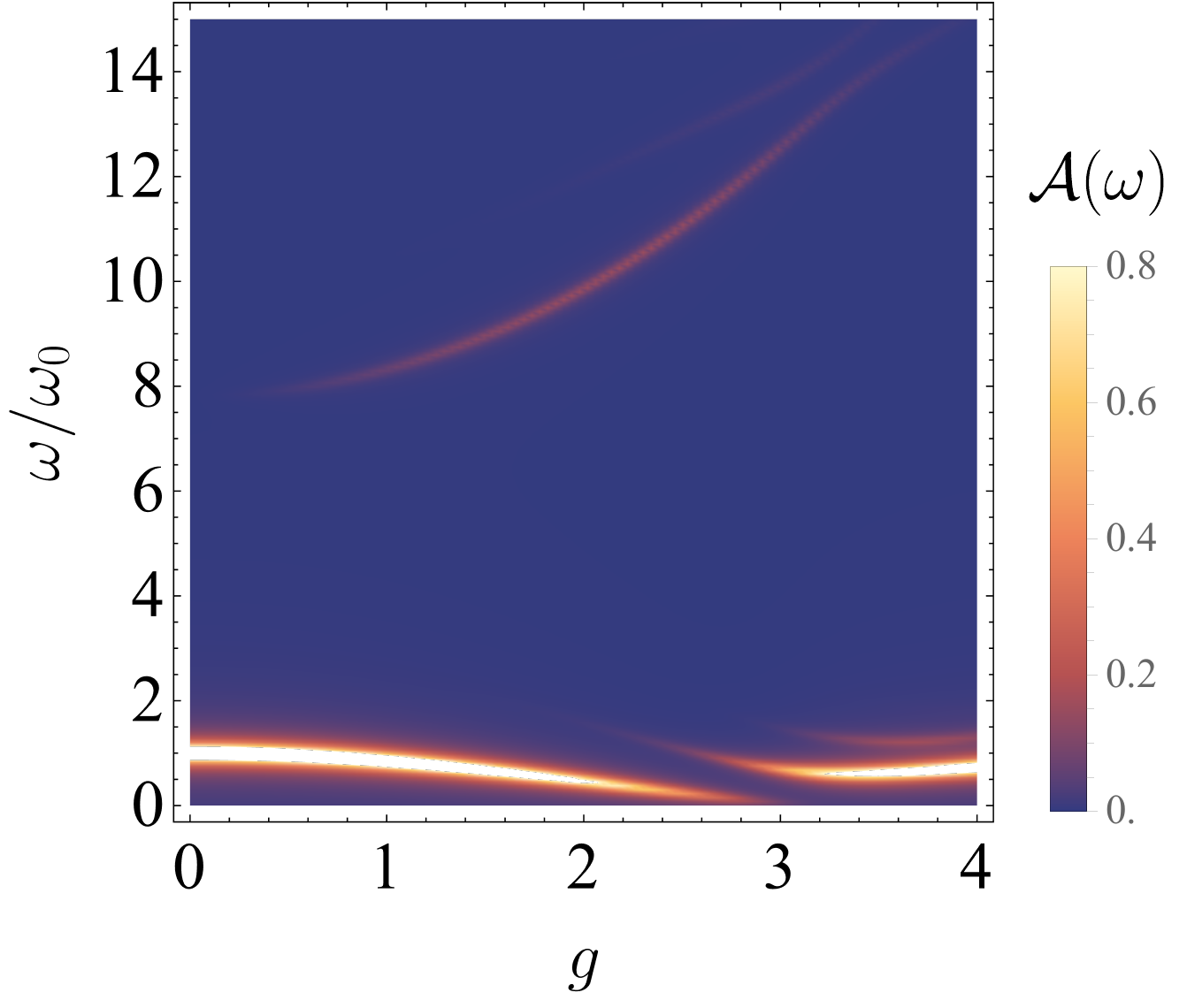}
\put(-175,140){(b)}
\caption{\label{fig:optical_cond_2nd_order_QPT} Density plot of (a) the optical conductivity $\operatorname{Re}\left[\sigma(\omega)\right]/\sigma_0$ and (b) the absorption spectrum (photon spectral function $\mathcal{A}(\omega)$) for $N = 2$ DQDs at $U/\omega_0=-5$ and $\omega_0/\Delta=0.1$ {corresponding to the smooth transition region}. The optical conductivity shows a frequency comb at strong coupling. The normalization parameter $\sigma_0 = (eb)^2$ and each delta peak is replaced by a Lorentzian with the broadening $\Gamma/\omega_0=0.1$. The absorption spectrum shows the lower polariton softening to zero {in the smooth transition region} alongside a reduction in its spectral weight. In the case of the first-order QPT, the photon spectral function shows a jump instead of softening to zero {(see Appendix~\ref{app_absorp_1st})}. {At $g=0$, the polariton frequencies are $\omega_0$ and the eigenvalues of $-\Delta S_x+U S_z^2$.}}
\end{figure}

Two gauge invariant response functions that can be routinely measured are the optical conductivity $\sigma(\omega)$
and the absorption spectrum. The latter is defined~\cite{louisell1973quantum} as the cavity response to an AC voltage applied to the cavity and {is proportional to $\mathcal{A}(\omega)=-2\operatorname{Im} G^{\text{R}}_{\text{ph}}(\omega)$, with $G^{\text{R}}_{\text{ph}}(t)=-i\theta(t)\langle[{A}_z(t),{A}_z(0)]\rangle$}, $\theta(t)$ being the Heaviside step function. The absorption spectrum was thoroughly studied before~\cite{PhysRevLett.102.186402,PhysRevX.4.041031,PhysRevLett.105.196402} as a function of the driving frequency, showing two standard polariton branches. We plot the absorption spectrum in Fig.~\ref{fig:optical_cond_2nd_order_QPT}(b), revealing the softening of the lower polariton mode to zero at the smooth transition.

The optical conductivity is calculated by standard means~\cite{mahan1990many} 
\begin{equation}
    \operatorname{Re}\left[\sigma(\omega)\right] = \frac{1}{2\omega}\int_{-\infty}^{+\infty}dt e^{i\omega t}\langle{J}(t){J}\rangle,
\end{equation}
where the current operator ${J}$ along the DQD axis is defined as ${J} = e \, d{z}/dt = i \left[{H}_D, (eb/2) \sum_{i}{d}_{i,z} \right]$ because the ${z}$ coordinate operator is replaced by the dipole moment, and in the length gauge [Eq.(\ref{HD})] is given by $J = i \left[H_D, eb S_z \right] = - e b \Delta S_y $.

Contrary to absorption, the optical conductivity retains a strong frequency comb~\cite{PhysRevLett.130.023601, Mehta2023, Eckhardt2022} deep in the ferroelectric phase, see Fig.~\ref{fig:optical_cond_2nd_order_QPT}(a).
We stress that such a frequency comb is not present in the semiclassical approximation, see Eq.~(\ref{meanfield}).
Here, we present an analytic result for $\operatorname{Re}\left[\sigma(\omega)\right]$,
\begin{eqnarray}
    && \operatorname{Re}\left[\sigma(\omega)\right] \approx \frac{\pi S}{2 \omega} \left(e b \Delta \right)^2 \sum\limits_{n = 0}^\infty p_n(g^2) \delta \left(\omega - E_n \right) , \label{siganalytic}
\end{eqnarray}
where $S$ is the pseudospin, $p_n(z) = e^{-z} z^n/n!$ is the Poisson distribution, and $E_n = n \omega_0 - U (2 S - 1)$.
Equation~(\ref{siganalytic}) is valid in the ferroelectric phase at arbitrary $g$ and near-full semiclassical polarization $|\langle S_z \rangle| \approx S$, {see Appendix~\ref{app_opt_cond} for details}.

Our findings are relevant for state-of-the-art experiments, providing key parameters: $\omega_0$, ranges from tens of GHz to THz; the splitting in DQDs can vary between $\Delta \sim 0.1-10$ meV; the Coulomb interaction, $|U|$, may reach several meV depending on the dot configuration. The light-matter coupling $g=\sqrt{W/\omega_0}$ is widely tunable and can significantly exceed unity if the length of each DQD is large and the mode volume is highly compressed~\cite{ModeCompressionNanoLett}.

\section{Conclusion} 
    We analyzed two DQDs coupled to a cavity mode and found a ferroelectric QPT at strong light-matter coupling and attractive dipole-dipole interaction between DQDs due to the Coulomb force. 
    {There is a first-order QPT and a smooth transition separated by a critical point.  We showed that the ground and the first excited states of two cavity-coupled DQDs in the smooth transition region are cat states protected by a finite energy splitting.
    We argue that such cavity-coupled DQD systems can be used as cat qubits. The quantum phase transition and the cat states are shown to persist against  cavity losses and variation of system parameters.} Higher excited states are studied via the optical conductivity which exhibits a frequency comb at strong coupling.

\section*{Acknowledgments}
This work was supported by the Georg H. Endress Foundation (VKK and DM)
and the Swiss National Science Foundation.
This project has received funding from the European
Union's Horizon 2020 research and innovation program
under Grant Agreement No 862046 and under
Grant Agreement No 757725 (the ERC Starting Grant). This work was supported as a part of NCCR SPIN, a National Centre of Competence (or Excellence) in Research, funded by the Swiss National Science Foundation (Grant No. 51NF40-180604).

\appendix

\section{Hamiltonian of the electronic system}
\label{dipole_int_deriv}
Here we derive the dipole-dipole interaction term introduced in the electronic Hamiltonian $H_{\text{el}}$ in the main text. 
We consider $N$ singly-occupied DQDs that interact with each other via the Coulomb repulsion, 
\begin{align}
&{H}_{\text{el}} = -\sum_{k=0}^{N-1} \left(\frac{\Delta}{2} {c}_{2k+1}^\dagger {c}_{2k+2} + \text{h.c.}\right)\nonumber\\
&+\frac{1}{2}\sum_{k\neq k'} W_{kk'} (n_{k}-n_{k,0}) (n_{k'}-n_{k',0}) \nonumber\\
&- \frac{V_b}{2} \sum_{k=0}^{N-1} (n_{2k+2}-n_{2k+1}), \label{hamel}
\end{align}
where the $2N$ sites comprising $N$ DQDs are located at the positions $z_k$ ($k = 1, \dots ,2N$) and numbered continuously, i.e. $c_{1}=c_{1,L}$, $c_{2}=c_{1,R}$, $c_{3}=c_{2,L}$, $c_{4}=c_{2,R}, \dots$, where ${c}_{i,L/R}$ are the electron annihilation operators introduced in the main text, $\Delta/2$ is the DQD level hybridization (the hopping amplitude), $V_b$ the bias in each DQD, $b$ the DQD length, i.e. $z_2-z_1 = z_4-z_3 =...=b$, $n_{k,0} = \langle \text{GS}_0|n_k|\text{GS}_0\rangle$ is the average occupation of the $k^{\mathrm{th}}$ site, $|\text{GS}_0\rangle$ is the ground state of the non-interacting Hamiltonian. The sum in the kinetic energy is restricted to odd numbers only as there is no hopping between the DQDs. 
The Coulomb interaction is described by $W_{kk'} =e^2/(\varepsilon|z_k-z_{k'} |)$, $\varepsilon$ is the dielectric constant. 
As we assume that each DQD is singly-occupied, there are no other interaction terms.

If there are only $N = 2$ singly-occupied DQDs, then $n_1 + n_2 = n_3 + n_4 = 1$ and the inter-DQD Coulomb interaction can be represented in terms of the product $n_2 n_3$ only. 
On the other hand, the dipole-dipole interaction $2U[(n_1-n_2)/2][(n_3-n_4)/2]$ from the main text can be simplified to $(U/2)(1-2n_2)(2n_3-1) = -2 U n_2 n_3 +U(n_2+n_3)-U/2$.
Comparing the coefficients of the bilinear term $n_2 n_3$ in the Coulomb term and in the dipole-dipole interaction term, we find the dipole-dipole interaction strength,
\begin{align}
&U = -\frac{1}{2}(-W_{13}+W_{14}+W_{23}-W_{24})\nonumber\\
&=-\frac{e^2}{2\varepsilon}(-\frac{1}{l+b}+\frac{1}{l+2b}+\frac{1}{l}-\frac{1}{l+b})\nonumber\\
&=-\frac{e^2}{\varepsilon}\frac{b^2}{l(l+b)(l+2b)} ,
\end{align}
where $l=z_3-z_2$.
The dipole-dipole interaction in case of arbitrary $N \ge 2$ is derived in a similar fashion.
Therefore, the Hamiltonian in Eq.~(\ref{hamel}) is equivalent to $H_{\text{el}}$ in the main text 
(up to a constant energy shift). 
The distance $z_3-z_2 = l$ is related to the distance between the DQD centers $r_{12}$ from the main text: $r_{12}=l+b$. 
At $l\gg b$ we restore the result from the main text $U=-(eb)^2/[\varepsilon r_{12}^3]$. 
The dipole-dipole approximation is exact due to the two-level truncation of DQD energy levels.

\section{Semiclassical analysis}
\label{semiclass_an}
The semiclassical Hamiltonian $H_\text{sm}$ takes the following form
\begin{align}
    &H_\text{sm} = \omega_0 a^\dagger a - V_b S_z -\Delta S_x + U S_z^2 \nonumber\\
    &+ g^2 \omega_0 \left(\delta S_z\right)^2,
\end{align}
where $\delta S_z = S_z - \langle S_z \rangle$, $\langle S_z\rangle$ is the average of $S_z$ over the ground state of $H_\text{sm}$.
Within this approximation, photons are decoupled from the orbital pseudospin, so the semiclassical ground state wave function $\Psi_\text{sm} = |0\rangle \chi_\text{sm}$, where $|0\rangle$ is the photon vacuum, $\chi_\text{sm}$ is the lowest-energy spinor of $\langle 0|H_\text{sm}|0\rangle$.
The semiclassical ground-state energy $E_\text{sm}$ follows from the characteristic equation $\det(\langle 0|H_\text{sm}|0\rangle - E_\text{sm}) = 0$.
\begin{figure}[b]
\includegraphics[width=0.8\columnwidth]{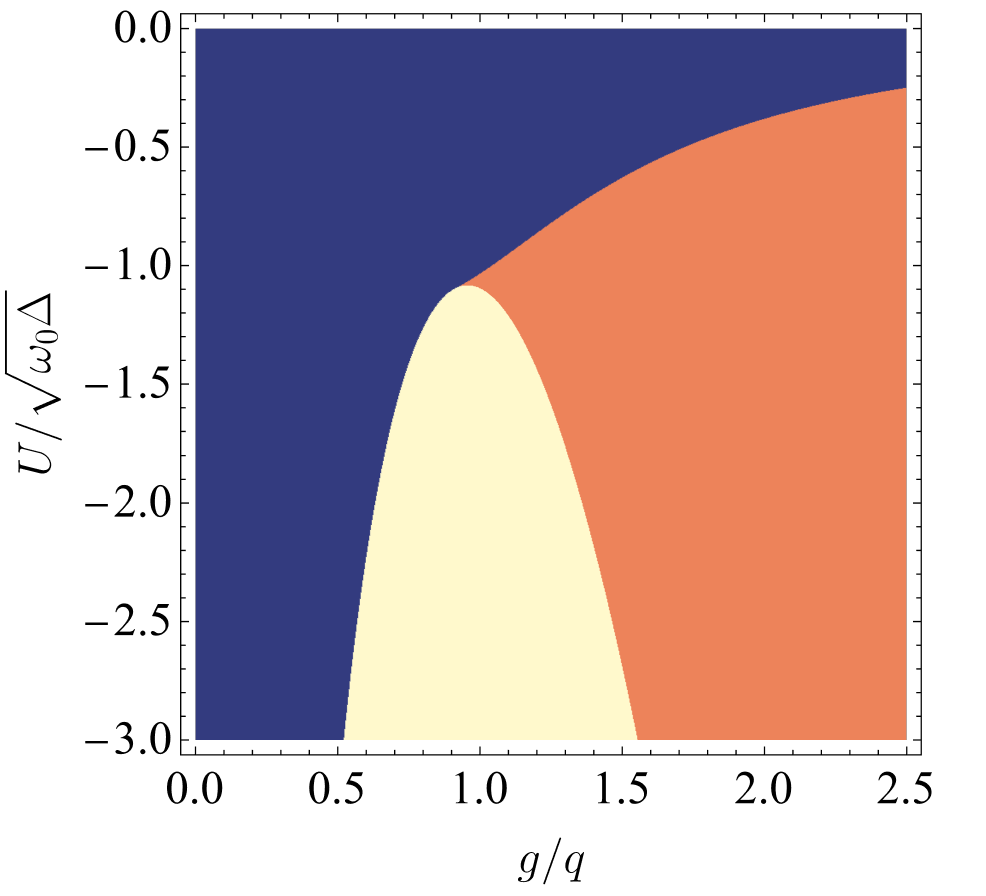}
\caption{\label{fig:phase_distinguisher} 
Energy landscape of the semiclassical ground state at $\omega_0/\Delta= 0.1$. 
Blue region: the semiclassical lowest energy $E_\text{sm}(\alpha)$ has its minimum at $\alpha=0$. Yellow region: the semiclassical energy $E_\text{sm}(\alpha)$ has the two minima $\pm\alpha\neq 0$, and $E^{\prime\prime}_\text{sm}(\alpha=0)<0$. Orange region: the semiclassical energy has the two minima at $\pm\alpha\neq 0$, and $E^{\prime\prime}_\text{sm}(\alpha=0)>0$. The transition from blue to the yellow region is a second-order quantum phase transition (QPT), whereas from blue to orange it is a 1st-order QPT. The boundary between yellow and orange regions does not correspond to a phase transition as in both regions the global minima are at $\pm\alpha\neq0$ (ferroelectric phase). {We note here that the second order QPT predicted by the semiclassical analysis turns into a smooth transition in the exact treatment of the problem.}}
\vspace{-5pt}
\end{figure}
In case $S = 1$, the characteristic equation is a third-degree polynomial. We introduced the notation $\alpha = g \langle S_z \rangle$, and we chose to measure all energies in $\sqrt{\omega_0\Delta}$. Then, all three roots of this characteristic equation are real and can be conveniently expressed via the dimensionless parameters $(g/q,U/\sqrt{\omega_0\Delta})$ (with  $q=\sqrt{\Delta/\omega_0}$) as follows:
\begin{align}
&\frac{E_k}{q\sqrt{\omega_0\Delta}} = \left(\frac{\alpha}{q}\right)^2 + \frac{2}{3} \left(\frac{U}{\sqrt{\omega_0\Delta}}+  \left(\frac{g}{q}\right)^2 \right) \nonumber\\
&- \frac{2}{3} \sqrt{\frac{P}{q^2 \omega_0\Delta}} \cos \left[\frac{1}{3}\arccos\left(\frac{Q}{P^{3/2}}\right) + \frac{2 \pi k}{3}\right] ,
\end{align}
where $k \in \{0, \pm 1\}$, and $P$ and $Q$ are given by 
\begin{align}
&\frac{P}{q^2 \omega_0\Delta} =\\
&=\left(\frac{U}{\sqrt{\omega_0\Delta}} +  \left(\frac{g}{q}\right)^2  \right)^2 +12\left(\frac{\alpha}{q}\right)^2\left(\frac{g}{q}\right)^2 + 3 ,\nonumber\\
&\frac{Q}{(q^2\omega_0\Delta)^{3/2}} = \left(\frac{U}{\sqrt{\omega_0\Delta}} + \left(\frac{g}{q}\right)^2\right) \\
&\cdot \left[ \left(\frac{U}{\sqrt{\omega_0\Delta}} + \left(\frac{g}{q}\right)^2 \right)^2 - 36\left(\frac{\alpha}{q}\right)^2\left(\frac{g}{q}\right)^2+\frac{9}{2}\right]\nonumber. 
\end{align}
The ground state corresponds to $k = 0$, i.e. $E_\text{sm} = E_{k = 0}$.

Considering $\alpha$ as a variational parameter, we analyze the global minima of $E_\text{sm} (\alpha)$ at all other parameters fixed.
We stress that $\alpha = g \langle S_z \rangle$ at extrema of $E_\text{sm} (\alpha)$.
In Fig.~\ref{fig:phase_distinguisher} we display three different regions: one global minimum (blue), two global minima located at $\pm \alpha \ne 0$ with $E''_\text{sm}(\alpha = 0) < 0$ (yellow) or with $E''_\text{sm}(\alpha = 0) > 0$ (orange).
Notice that $E_\text{sm} (\alpha)$ contains two (three) local minima in the yellow (orange) region.
{In other words, in the semiclassical analysis the boundary between blue and yellow (blue and orange) regions corresponds to the second- (first-) order ferroelectric quantum phase transition (FPT, we use the terms FPT and QPT interchangeably in this work).}
The boundary between yellow and orange regions does not correspond to a phase transition, it only shows that the local extremum at $\alpha = 0$ changes from local maximum to local minimum, while the global minima are located at $\pm \alpha \ne 0$. The position of the critical point separating the first-order QPT from the smooth transition remains unchanged if plotted in coordinates ($g/q$, $U/\sqrt{\omega_0\Delta}$) when the quasi-thermodynamic limit $q=\sqrt{\Delta/\omega_0}\to\infty$ is considered.

\section{Non-equivalent quantum dots}
\label{non_equiv_QD}
\begin{figure}[h]
	\includegraphics[width=0.74\columnwidth]{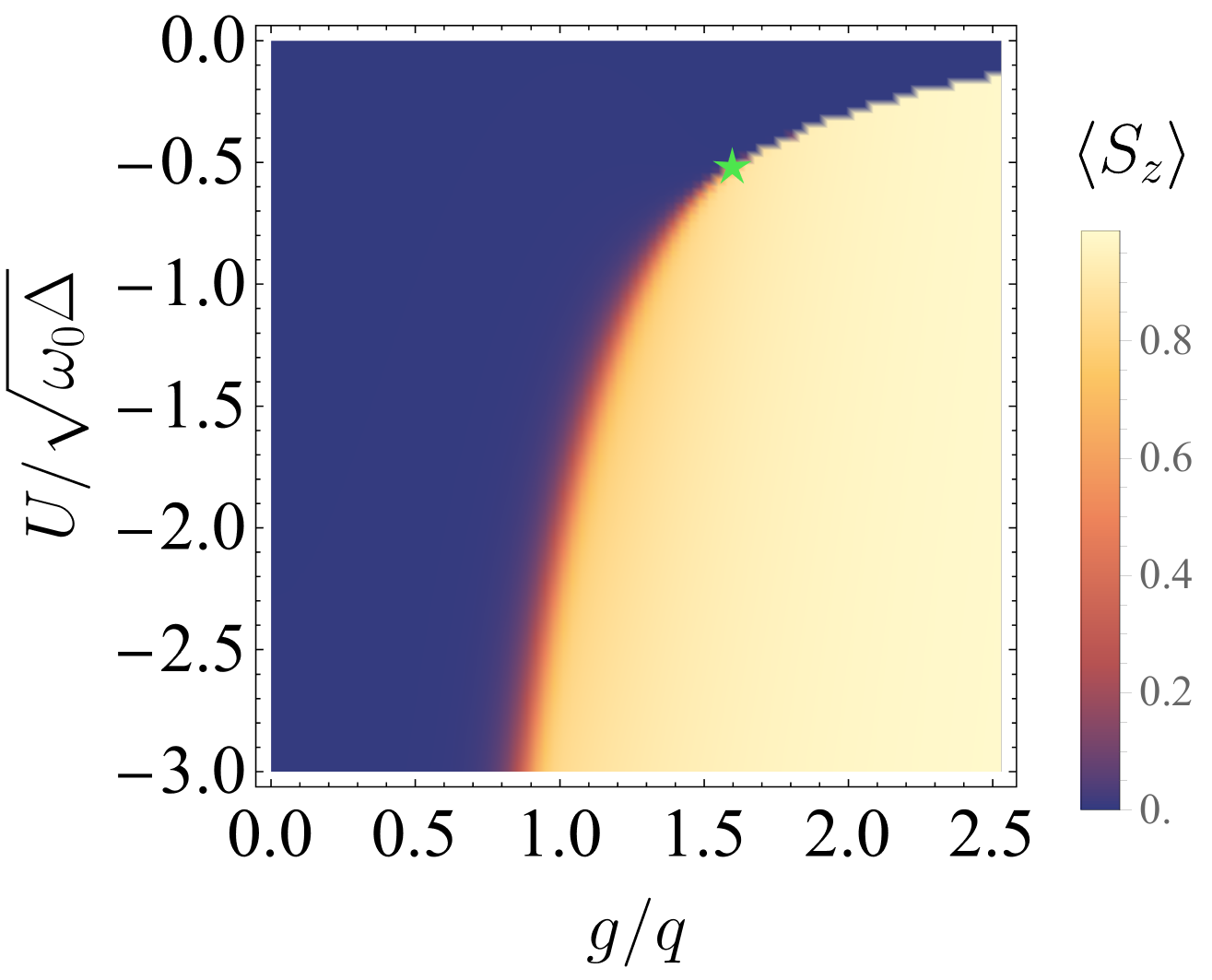}\put(-182,137){(a)}\\
	 \includegraphics[width=0.74\columnwidth]{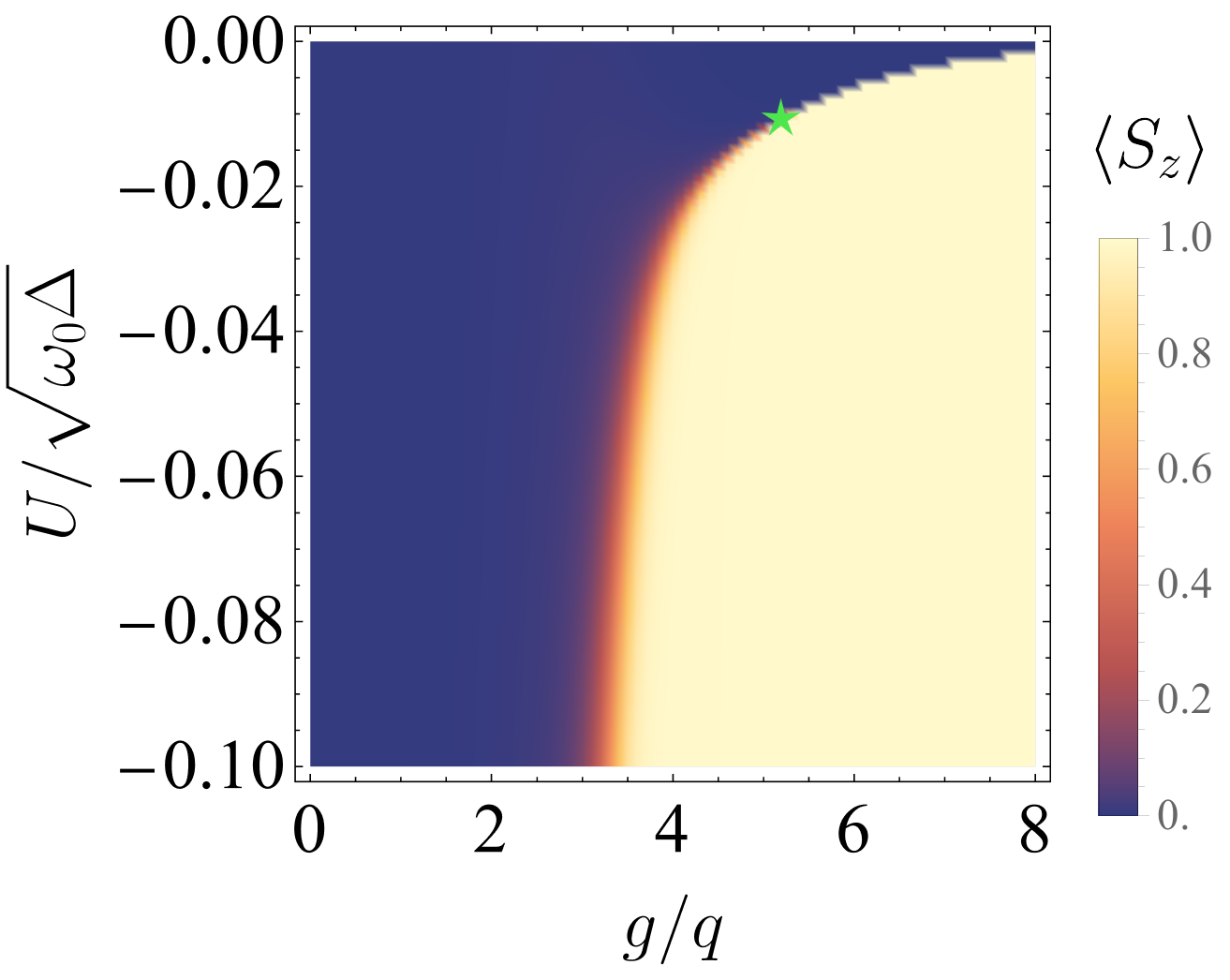}\put(-182,135){(b)}
	\caption{\label{fig:Sz_map_non_ideal}  The map of the net dipole moment $\langle S_z\rangle$ of the non-ideal system at  $g_2/g_1=0.8$ ($g_1=g$), $\Delta_2/\Delta_1=1.05$ and $V_{b,1}/\sqrt{\omega_0\Delta}=10^{-3}$,  $V_{b,2}/V_{b,1}=0.8$. The star marks the critical point separating the first- and second-order quantum phase transitions.
	(a): $\omega_0/\Delta_1=1/q^2=0.1$. (b): $\omega_0/\Delta_1 =1/q^2= 1$.}
	\vspace{-5pt}
\end{figure}

If the DQDs are not equivalent, i.e. have different splittings $\Delta_i$, applied biases $V_{b,i}$, widths $b_i$ (and, hence, couplings to the cavity $g_i$), the model describing a set of $N=2$ DQDs placed in the cavity from the main text takes the following form
\begin{align}
 & {H}_0 = \omega_0 {a}^\dagger {a} - \frac{1}{2}\sum_{i=1}^{N=2}V_{b,i} {\sigma}_{i,z} + U {S}_z^2 , \label{H0_app} \\
 &{\tilde{H}} = {H}_0 \nonumber\\
 &- \frac{1}{2}\sum_{i=1}^{N=2}\frac{\Delta_i}{2} \left(e^{g_i (a - a^\dagger)} {\sigma}_{i,+} + e^{-g_i (a - a^\dagger)} {\sigma}_{i,-} \right)  ,\label{eq:Peierls_Hamilt_non_ideal}\\
 & g_i = \sqrt{\frac{W_i} {\omega_0}} \, , \hspace{5pt} W =   \frac{2 \pi e^2 b_i^2}{\varepsilon V_\text{eff}}. \label{g} 
\end{align}
We note that in the present case the Coulomb term $U {S}_z^2$ retains its form and only the expression of $U$ via microscopic characteristics of the individual DQDs is altered. {It is clear from Fig.~\ref{fig:Sz_map_non_ideal} that  the first-order phase boundary remains sharp both near the quasi-thermodynamic limit and away from it.}

\section{Phase diagrams: exact diagonalization vs semiclassics}
\label{semiclass_phase_diag}
\begin{figure}[h]
		\includegraphics[width=0.76\columnwidth]{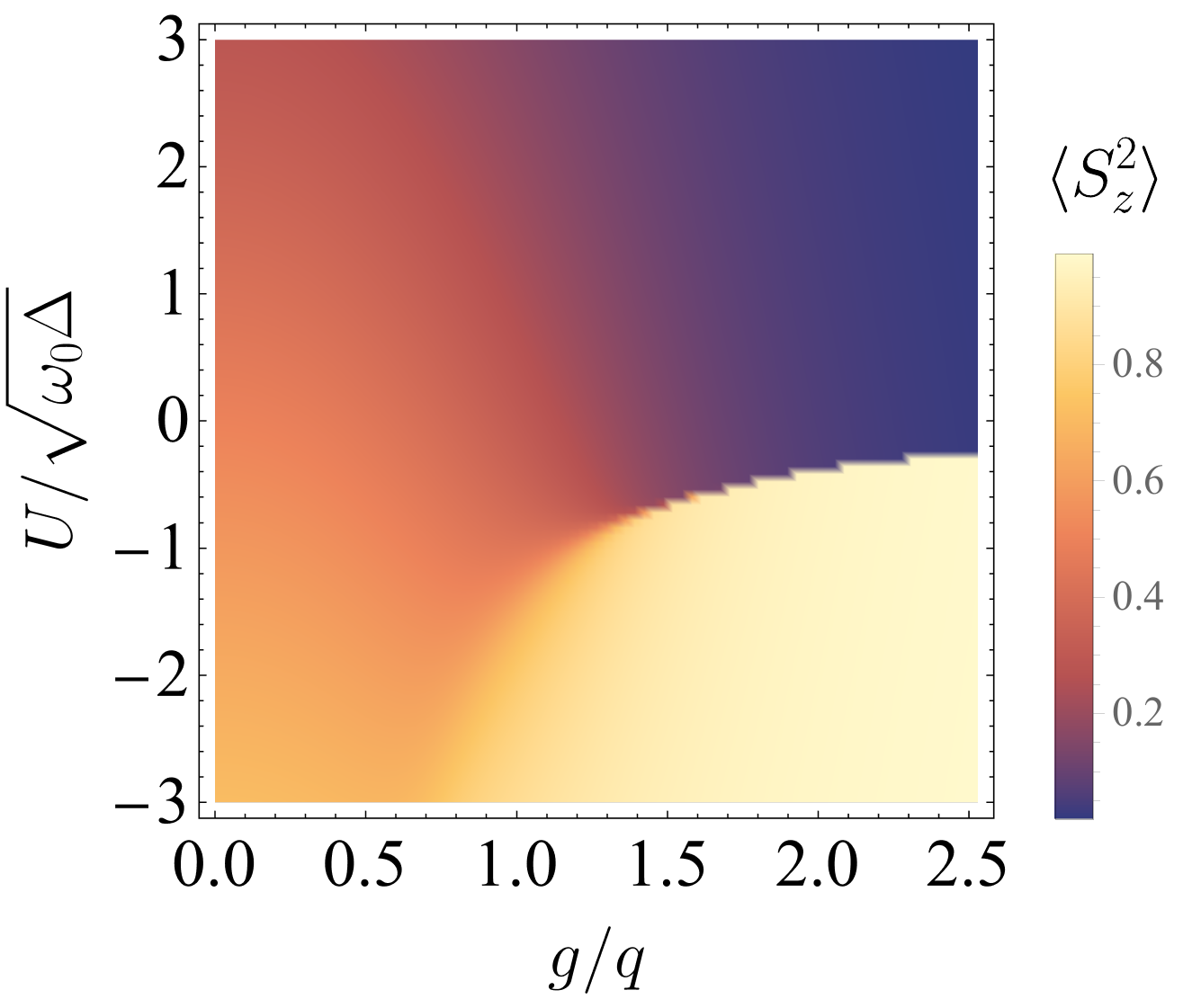}\put(-185,150){(a)}\\
		\includegraphics[width=0.76\columnwidth]{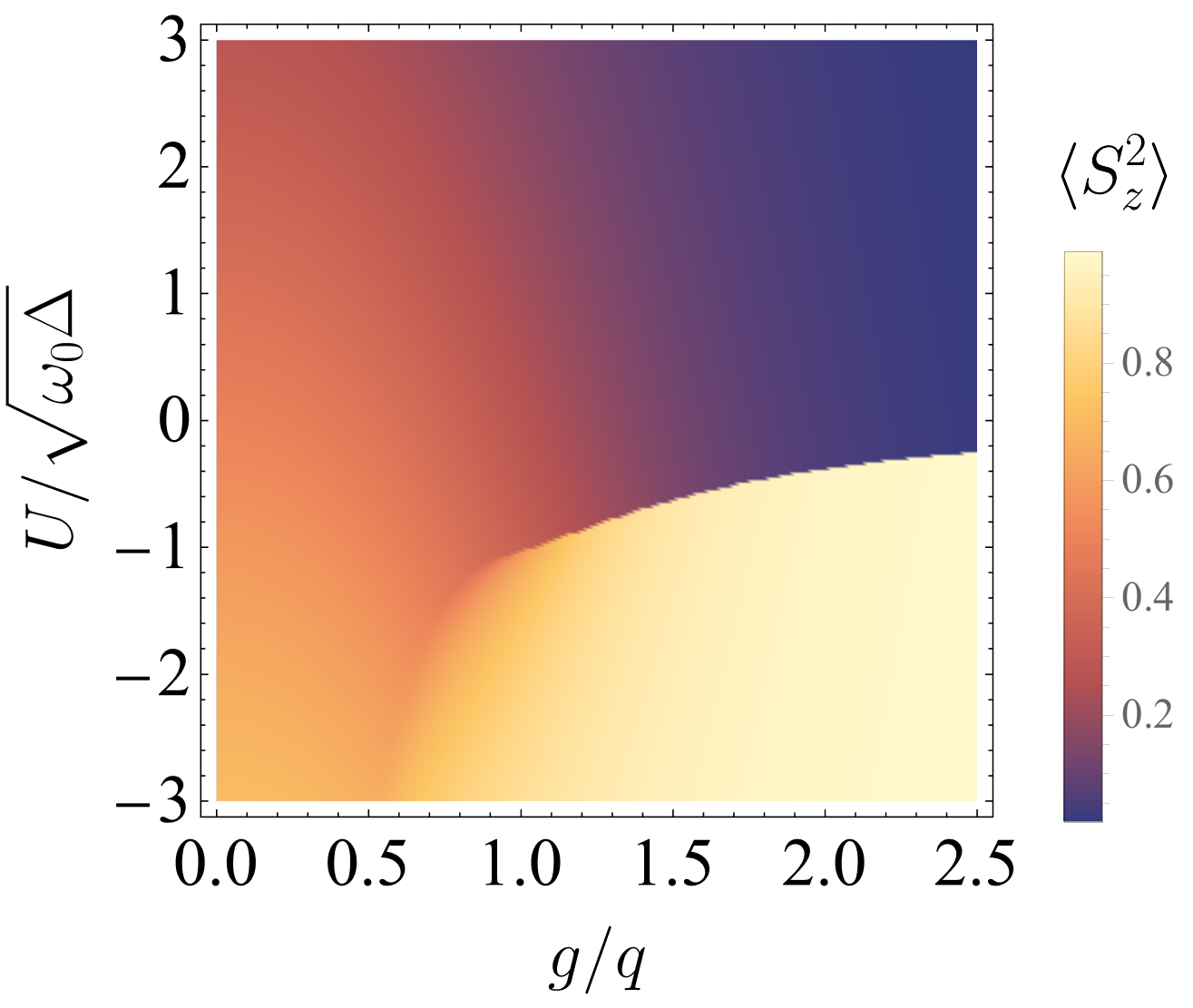}\put(-185,150){(b)}
	\caption{\label{fig:fig_SzSquared_maps} (a) Colour map of $\langle{S}_z^2\rangle$ obtained by the exact diagonalization. (b) Colour map of $\langle{S}_z^2\rangle$ obtained from the semiclassical solution, see the main text. In both figures $\omega_0/\Delta=0.1$. 
	The phase boundary predicted by the semiclassical approximation approaches the exact one in the quasi-thermodynamic limit $\omega_0/\Delta \to 0$. }
	\vspace{-5pt}
\end{figure}

In the case of a single DQD, the square of the dipole moment $S_z^2$ is trivial (identity matrix).
This is not the case for $N \ge 2$ DQDs. 
In Fig.~\ref{fig:fig_SzSquared_maps}(a),(b) we show the exact (numerical) and the semiclassical color maps of $\langle S_z^2 \rangle$ for two DQDs at $\omega_0 /\Delta = 0.1$ and zero temperature, $T = 0$.
Even though at $T = 0$ the phase boundaries on the $\langle S_z\rangle$ and $\langle S_z^2 \rangle$ color maps are the same, the situation is different at finite temperature $T \gg |V_b|$.
At these temperatures, $\langle S_z\rangle = 0$ due to the symmetry restoration effect, while $\langle S_z^2 \rangle$ is not sensitive to either weak symmetry breaking field $V_b \ll \omega_0$, or to the symmetry restoration due to the quantum tunneling (instanton) effect.
This is why we plot the $\langle S_z^2 \rangle$ color map at finite temperature in Fig.~\ref{fig:Sz_map}(b) in the main text.

\begin{figure}[h]
		\includegraphics[width=1\columnwidth]{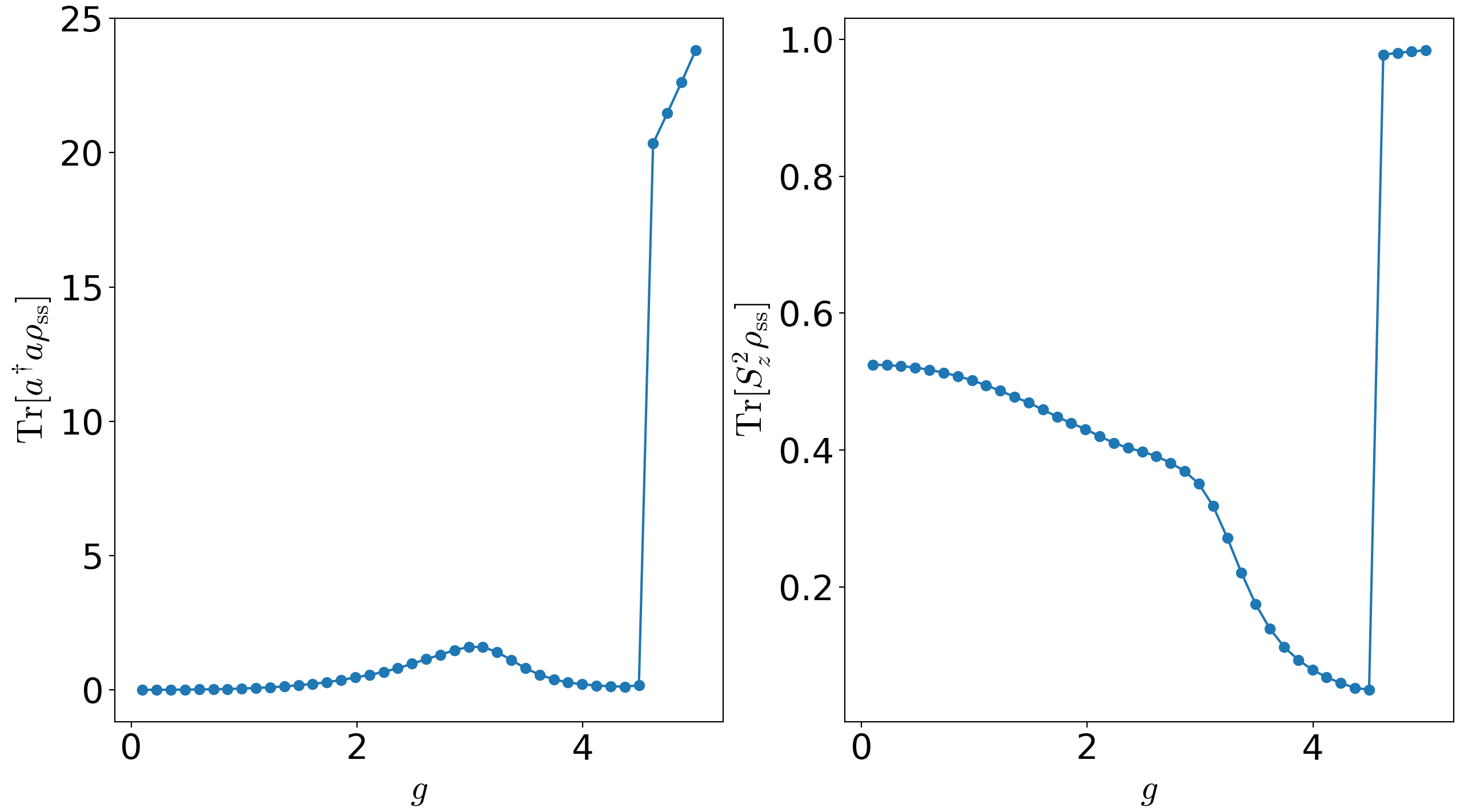}
		\put(-225,120){(a)}
		\put(-100,120){(b)}\\
		\includegraphics[width=1\columnwidth]{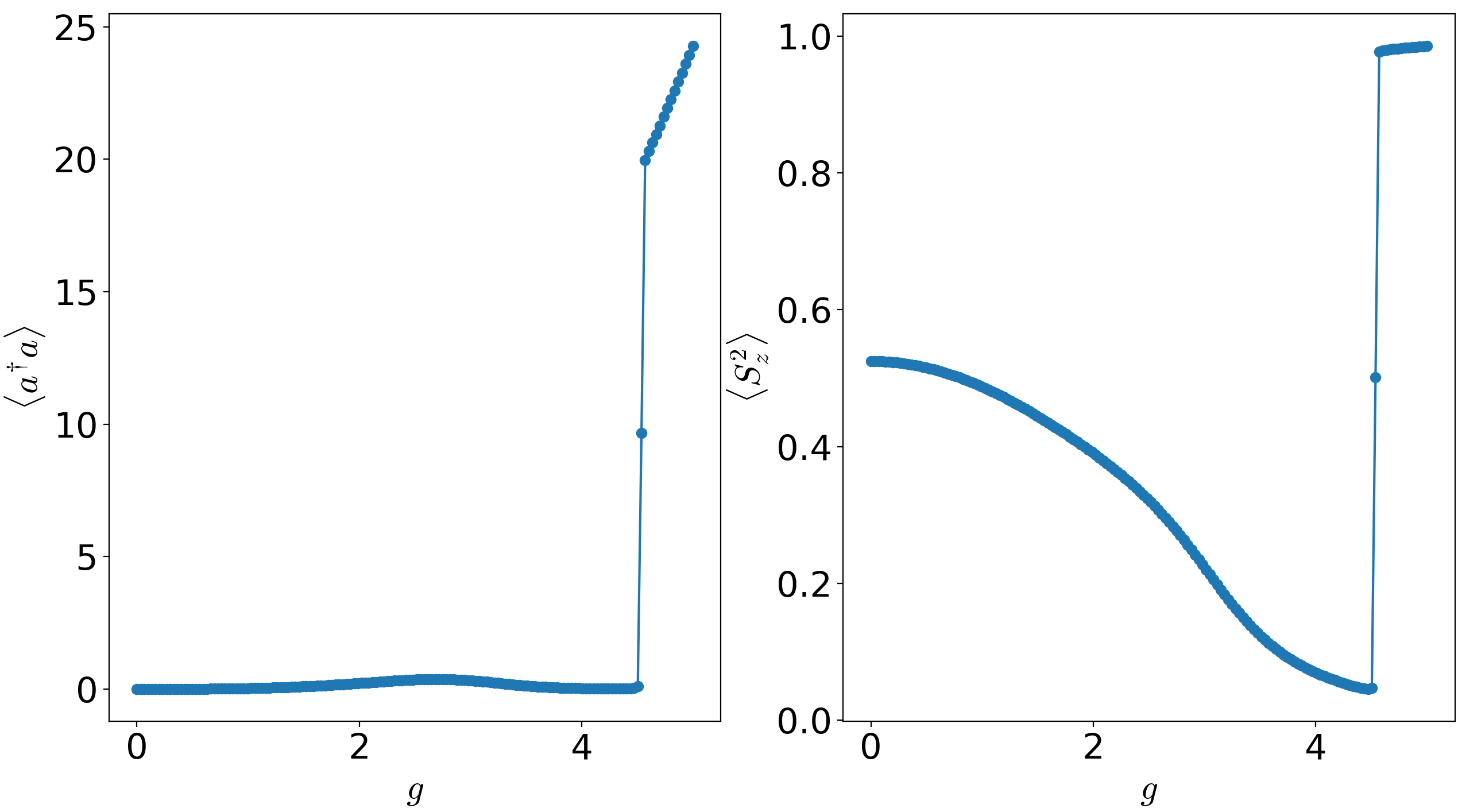}
		\put(-225,120){(c)}
		\put(-100,120){(d)}
	\caption{\label{fig:Lindblad} {(a,b): steady state solutions (photon number and the average value of the square of the net dipole operator, the later is a gauge invariant quantity) of the Lindblad equation with single-photon losses. (c,d): exact diagonalization of the closed system exhibiting a first-order quantum phase transition. The parameters are $U/\omega_0=-0.4$, $\Delta/\omega_0=4$, and $V_{b}/\omega_0\approx10^{-2}$.}}
	\vspace{-5pt}
\end{figure}

{
\section{Lossy cavity: the Lindblad equation}
\label{app_Lindblad}
In this section we show the solution of the Lindblad equation describing single-photon cavity losses:
\begin{equation}
\dot{\rho}=-i[H_{\text{l.g.}},\rho]+\gamma\mathcal{D}[a](\rho),
\end{equation}
where $\mathcal{D}[a](\rho)=a\rho a^\dagger-\frac{1}{2}(a^\dagger a\rho+\rho a^\dagger a)$. We use the length-gauge description with the Hamiltonian
\begin{align}
&{H}_{\text{l.g.}} = {\mathcal{U}} {H} {\mathcal{U}}^\dagger =\omega_0 {a}^\dagger {a} \\
&\nonumber-V_b{S}_z-{S}_x + U{S}_z^2 + g^2 \omega_0 \left( {S}_z\right)^2 - g \omega_0 \, {S}_z \left( {a} + {a}^\dagger\right) ,
\end{align}
where ${H}$ is given in the main text, and ${\mathcal{U}}=\exp{\left[g \, {S}_z({a}^\dagger-{a})\right]}$. The only difference between $H_{\text{l.g.}}$ and $H_{\text{D}}$ from the main text is that here we just performed the gauge transformation from the velocity to the length gauge without subtracting $\langle S_z\rangle$ in the unitary transformation. In Fig.~\ref{fig:Lindblad} we see that in the presence of single-photon losses in the cavity, the open system exhibits a 1st order quantum phase transition in the steady state $\rho_{ss}=\rho(t\to\infty)$ that is very similar to what the closed-system analysis from the main text predicts. Given that the numerical solution of the Lindblad equation requires higher truncation of the photon Hilbert space, we decided to choose $\Delta/\omega_0=4$, $U/\omega_0=-0.4$, a choice which also supports the 1st order QPT at a similar value of $g$ as in the main text.}

\section{Quantum jumps (Monte Carlo) analysis of the cat states}
\label{app_MC}
\begin{figure}[h]
		\includegraphics[width=0.49\columnwidth]{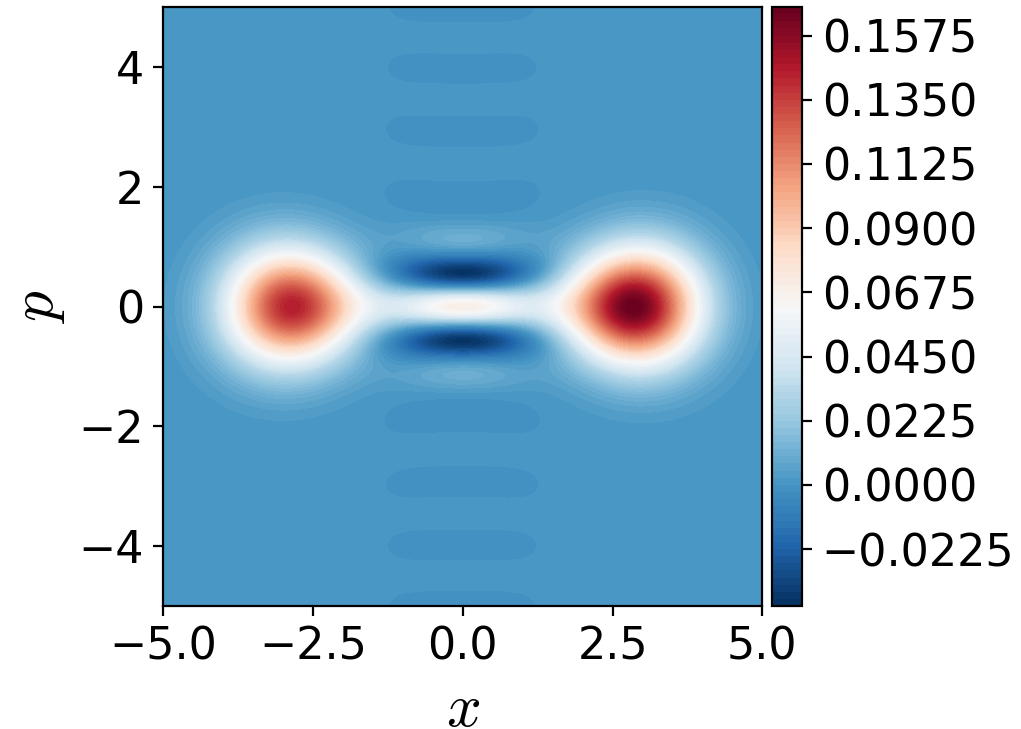} \put(-123,76){(a)} 
		\includegraphics[width=0.48\columnwidth]{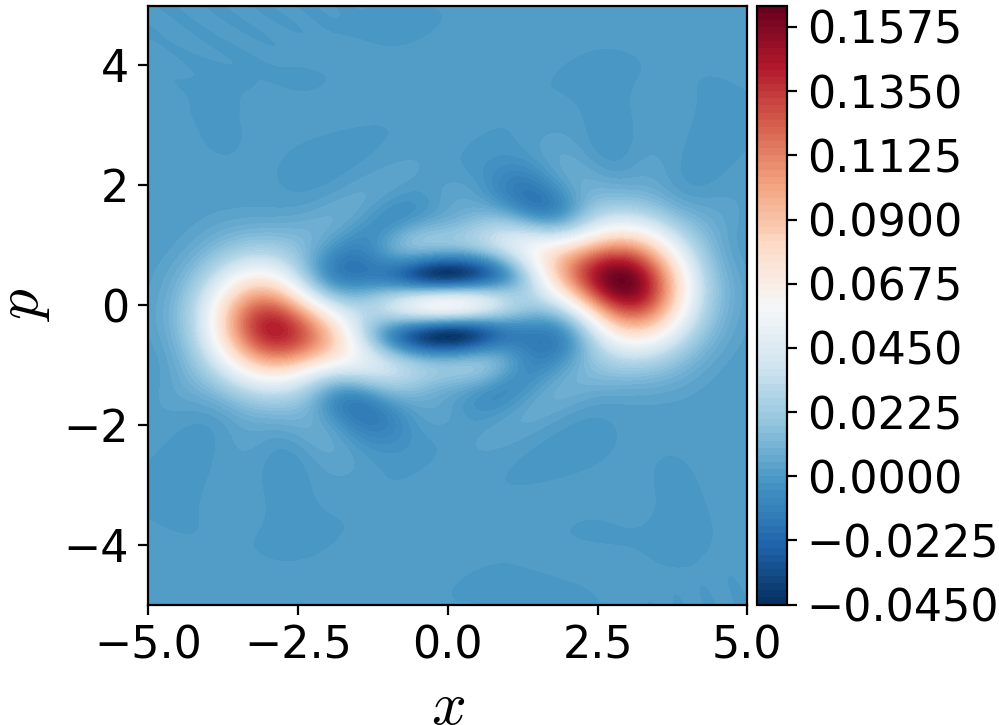}\put(-120,76){(b)}\\
		\includegraphics[width=0.97\columnwidth]{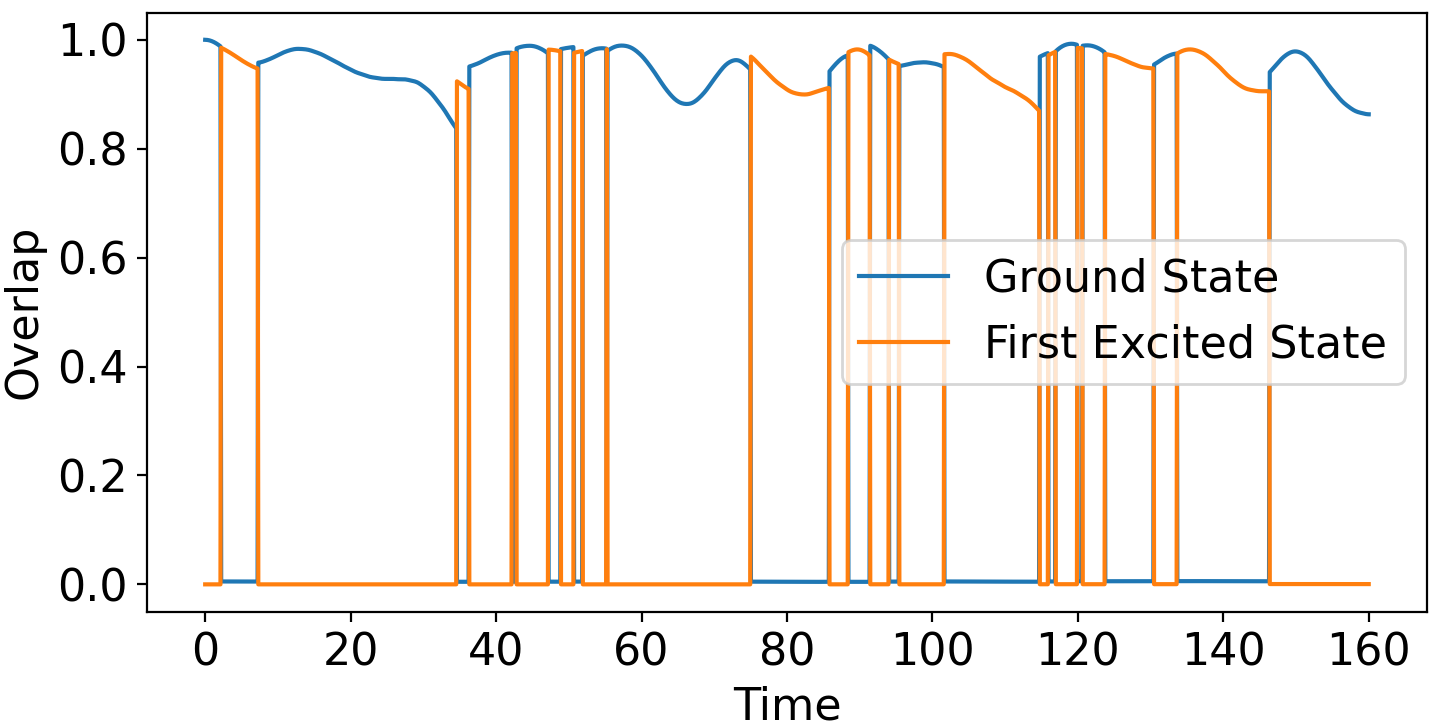}\put(-242,105){(c)}\\
		\includegraphics[width=0.58\columnwidth]{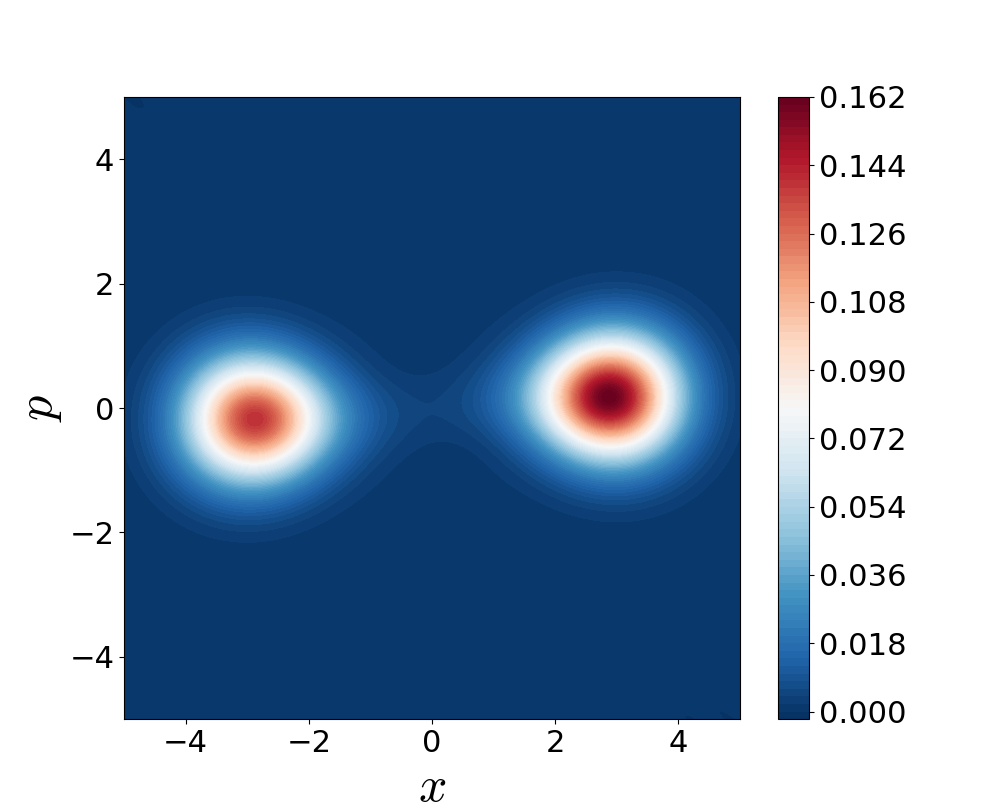}\put(-145,92){(d)}
	\caption{\label{fig:quant_traj}  {The Wigner function at the initial (a) and final (b) moments of the evolution of the lossy system initially prepared in the ground state at $g\approx2$, $U/\omega_0=-4$, $\Delta/\omega_0=4$, and $V_{b}/\omega_0\approx10^{-2}$. The Wigner function is calculated via the quantum Monte Carlo solver in Qutip~\cite{Johansson2012,Johansson2013} with $\gamma=0.05\omega_0$ ($1/\gamma \ll t_{\text{final}}$) and the truncation of the photon Hilbert space is set to $150$ photons. (c) The overlap between the current state in the quantum trajectory and the ground/first excited states as a function of time shows that the system throughout its evolution jumps between the ground and first excited states under the influence of the collapse operators. (d) The Wigner function in the steady state (the initial state is set to the ground (cat) state), calculated from the Lindblad master equation for the same parameters as in (a). This result corresponds to averaging over many quantum trajectories (with one particular realization given in (a)), showing the finiteness of the coherence time of the cat qubit in the presence of losses.}}
	\vspace{-5pt}
\end{figure}

Cat states are analyzed via the Wigner function defined as follows:
\begin{equation}
\mathcal{W}(q,p)=2\operatorname{Tr}\left[\rho D(q,p)\exp{(i\pi a^\dagger a)}\right],
\end{equation}
where $\rho$ is the cavity density matrix (i.e. the density matrix of the system with the electronic degrees of freedom traced out), the displacement operator can be expressed in terms of canonical coordinates $q$ and $p$,
\begin{equation}
D(q,p)=\exp{\left(\sqrt{2}\left(\left(q+i p\right)a^\dagger-\left(q-i p\right)a\right)\right)}.
\end{equation}
The Wigner function calculated in the length gauge is plotted in Fig.~\ref{fig:quant_traj}.

In order to demonstrate the stability of the cat states, we plot the Wigner function calculated in the length gauge for a lossy cavity with a symmetry-breaking field included. The analysis is performed with the help of the quantum jump method (Monte Carlo)~\cite{Carmichael1993}, which boils down to solving the Schr\"{o}dinger equation with the following non-Hermitian effective Hamiltonian
\begin{equation}
H_{\text{eff}}=H-\sum_n\frac{i}{2}C_n^\dagger C_n,
\end{equation}
where $C_n$ are collapse operators. In our case there is only one collapse operator $C=\sqrt{\gamma}a$ describing single-photon losses in the cavity. The main idea of the method is that one choses a random number $r\in[0,1)$ and propagates the state with the non-Hermitian Hamiltonian $H_{\text{eff}}$ until the moment of time $t=t_{\text{jump}}$ when $\langle\psi(t_{\text{jump}})|\psi(t_\text{jump})\rangle=r$. At this moment, the wave function undergoes a jump into a projected state using the collapse operator $C_n$ ($C_n$ is chosen with a relative probability of $\langle\psi(t_\text{jump})|C_n^\dagger C_n |\psi(t_\text{jump})\rangle$): $|\psi(t_\text{jump})\rangle\to C_n |\psi(t_\text{jump})\rangle/\langle\psi(t_\text{jump})|C_n^\dagger C_n|\psi(t_\text{jump})\rangle^{1/2}$, then a new random value of $r\in[0,1)$ is chosen and the propagation is continued. An individual realization is called a trajectory and below we show the numerically calculated trajectory in Fig.~\ref{fig:quant_traj}(a). It shows random switchings between two cat states on the typical time scale set by $\gamma^{-1}$, determining the coherence time of such a cat qubit. This behaviour can be understood by noting that $a|\Psi_{\text{G}}\rangle\propto|\Psi_{\text{E1}}\rangle$ and $a|\Psi_{\text{E1}}\rangle\propto|\Psi_{\text{G}}\rangle$, in other words single-photon losses introduce bit-flip errors. Averaging of many trajectories leads to agreement with the results obtained by the Lindblad approach (corresponding to averaging over an ensemble): from the behaviour of the trajectory it's clear that beyond the coherence time the interference fringe will average to zero (while the blobs remain) as shown in Fig.~\ref{fig:quant_traj}(b) due to the overlay of the even and odd cat states, however within the coherence time of such a cat qubit the system remains in the cat state.

\section{Photon spectral function at the 1st order QPT}
\label{app_absorp_1st}
\begin{figure}[h]
		\includegraphics[width=0.7\columnwidth]{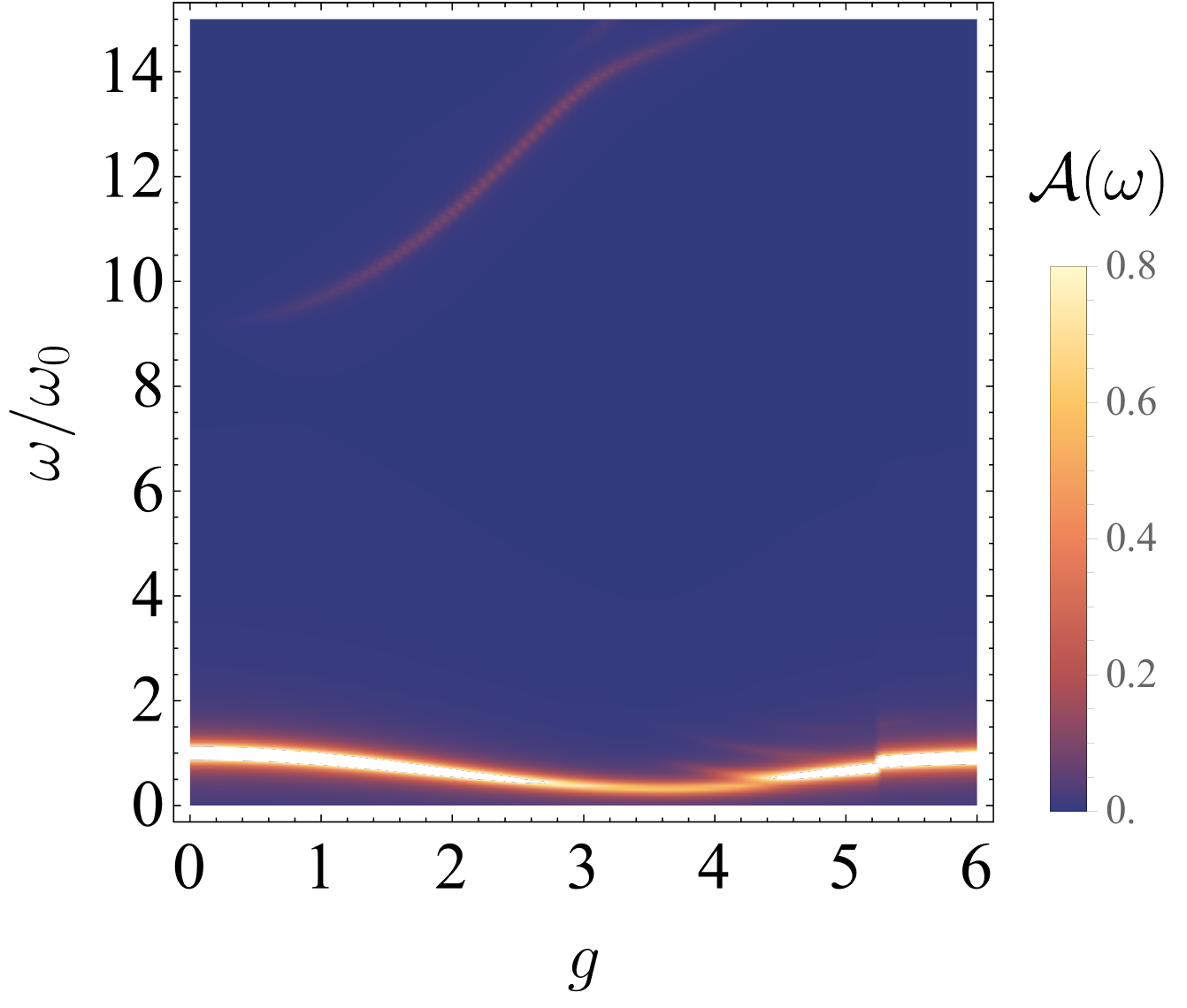}\put(-176,137){(a)}\\
		\includegraphics[width=0.7\columnwidth]{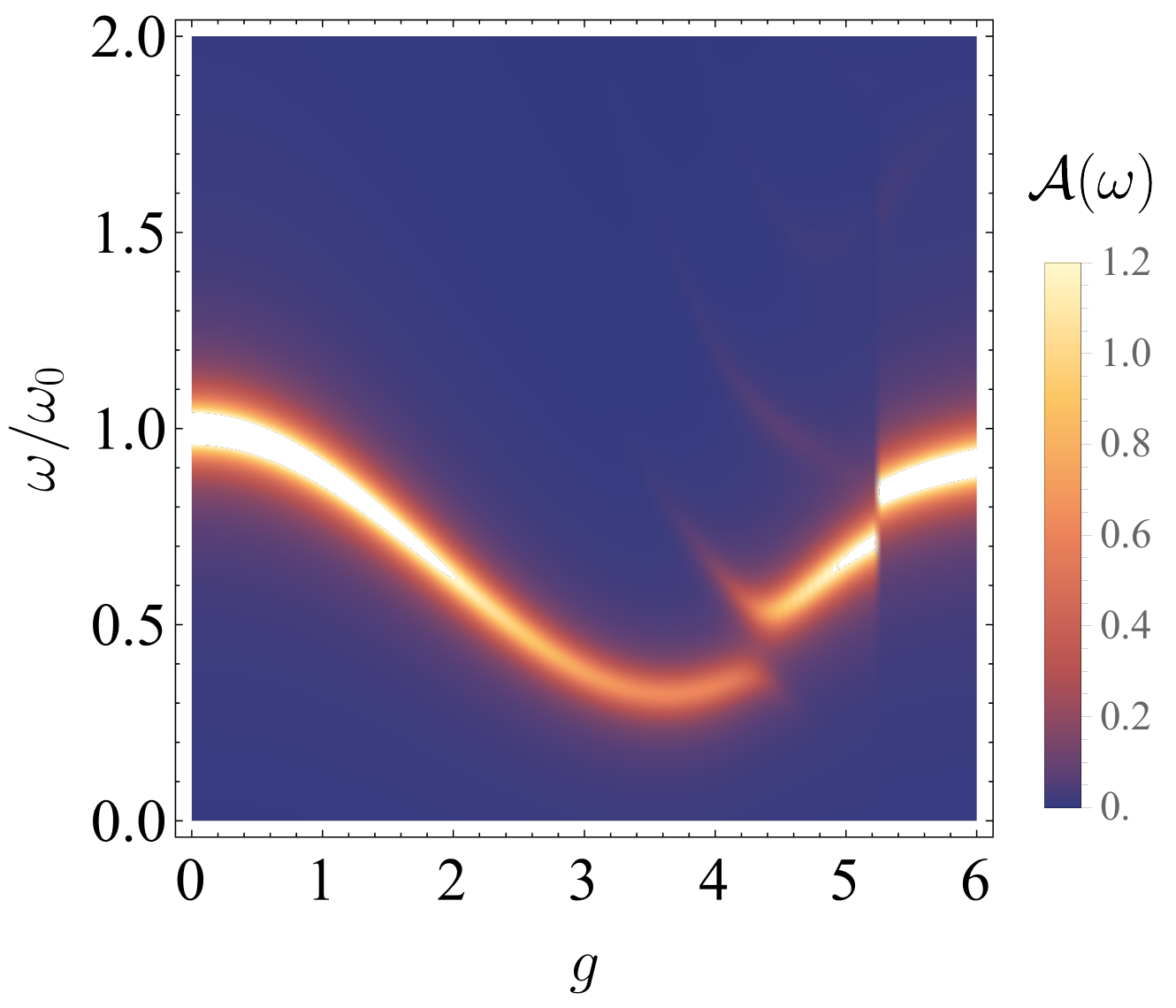}\put(-176,137){(b)}
	\caption{\label{fig:polariton_spectrum_1st_order_QPT} (a) Density plot of the absorption spectrum (photon spectral function) for $N = 2$ DQDs at $U/\omega_0=-1.75$ and $\omega_0/\Delta=0.1$. The absorption spectrum shows a discontinuity at the first-order QPT.  The panel (b) shows only the lower polariton branch at same parameters, clearly demonstrating a discontinuity at $g\approx 5.3$. Observed anticrossing at $g \approx 4.5$ is due to the replica polaritons.}
	\vspace{-5pt}
\end{figure}

In this section, we show that the absorption spectrum (the photon spectral function) demonstrates a sharp discontinuity at the first order QPT, see Fig.~\ref{fig:polariton_spectrum_1st_order_QPT}. This discontinuity reflects a sudden jump of the net dipole moment on two sides of the first-order QPT.

\section{Optical conductivity in the ordered phase}
\label{app_opt_cond}
The optical conductivity $\sigma(\omega)$ is a gauge-invariant observable.
Here, we derive $\textrm{Re}[\sigma(\omega)]$ in the leading order in $\Delta$.
This result is applicable deep in the ordered phase where $\langle S_z \rangle \approx S$, $\langle S_z \rangle$ is the ground-state average of the semiclassical Hamiltonian $H_\text{sm}$, and $S$ is the value of orbital pseudospin.
Here, we assume $S \ge 1$, such that the dipole-dipole interaction $U S_z^2$ is non-trivial.
Indeed, in this case, effects of the ``depolarization'' field $-\Delta S_x$ are weak and therefore, they can be treated via the perturbation theory.
Note that the light-matter coupling constant $g$ can be of arbitrary value and the perturbation expansion is performed only in the small parameter $\Delta/|U S|$.
It is more convenient to present the derivation within the velocity gauge, see Eq.~(\ref{eq:Peierls_Hamilt}) in the main text.
The current operator $J$ along the DQD axis $z$ follows from the fact that the coordinate operator is given by $z = b S_z$ in the Peierls gauge, where $b$ is the separation between left and right minima within each DQD:
\begin{align}
    & J = e \frac{d z}{d t} = i \left[H, e b S_z \right] = \nonumber\\
    &i \frac{e b \Delta}{2} \left(e^{g (a - a^\dagger)} S_+ - e^{-g (a - a^\dagger)} S_- \right) . \label{J}
\end{align}
First, we calculate the current-current correlator $\Pi(g, t)$,
\begin{eqnarray}
    && \Pi(g, t) = - i \theta(t) \Bigl \langle \left[J^H (t), J(0)\right] \Bigr \rangle , \label{Pi}
\end{eqnarray}
where 
%$g$ is the light-matter coupling, 
$J^H(t) = e^{i H t} J e^{- i H t}$ is the Heisenberg representation of the current operator and $\theta(t)$ the Heaviside step function.
Within  leading order in $\Delta$, $\Pi(g, t)$ is given by the following average:
\begin{eqnarray}
    && \Pi_0(g, t) =  - i \theta(t) \Bigl \langle \left[J(t) , J(0) \right] \Bigr \rangle , \label{Pi0}
\end{eqnarray}
where $J(t) = e^{i H_0 t} {J} e^{-  i H_0 t}$ is the interaction representation of the current operator, $H_0$ is given by Eq.~(\ref{H0}) in the main text.
First we note that  $S_z(t) = S_z$ as $[H_0, S_z] = 0$.
The interaction representations of $S_+$ and $a$ are the following:
\begin{eqnarray}
    && S_+(t) = e^{-i K(S_z) t} S_+ = S_+ e^{- i K(S_z + 1) t} , \label{Splus} \\
    && a(t) = e^{- i \omega_0 t} a , \label{a}
\end{eqnarray}
where $K(S_z) = H_0 (S_z - 1) - H_0(S_z) = V_b + U (1 - 2 S_z)$.
The statistical average of the exponential operators then follows directly from the Campbell-Baker-Hausdorff formula,
\begin{align}
\label{F}
    &F(g, t) \equiv \Bigl \langle e^{ g ({a}(t) - {a}^\dagger (t))} e^{-  g ({a} - {a}^\dagger)} \Bigr \rangle\\
    &= e^{- g^2 \left(2 N_\text{ph} + 1\right)}   \exp\left[2 g^2 N_\text{ph} \cos(\omega_0 t)  + g^2 e^{-i \omega_0 t}\right]\nonumber,
\end{align}
where $N_\text{ph} = [e^{\beta \omega_0}  - 1]^{-1}$ is the average photon number at finite temperature $T = 1 /\beta$.
As $\langle {S}_+(t) {S}_+ \rangle = \langle {S}_-(t) {S}_- \rangle  = 0$, 
we find
\begin{eqnarray}
    && \Pi_0(g, t) = F(g, t) \Pi_0(g = 0, t) , \label{Pi0factor}
\end{eqnarray}
where $ \Pi_0(g = 0, t)$ is the current-current correlator of the electron system decoupled from photons.
We emphasize that the factorization in Eq.~(\ref{Pi0factor}) holds in the limit $\Delta \ll \langle K({S}_z)\rangle$, i.e. when the hopping $\Delta$ can be treated as a small perturbation.
In the limit $\Delta = 0$, $H = H_0$, see Eq.~(\ref{H0}) in the main text, and the ground state at $U < 0$ is the state with the maximal pseudospin projection ($S_z = S$ at $V_b > 0$ and $S_z = -S$ at $V_b < 0$),
\begin{eqnarray}
&& \Pi_0(g = 0, t) = -i \theta(t) \frac{S}{2} \left(e b \Delta\right)^2 e^{- i t E_\text{opt}} , \label{Pifree}
\end{eqnarray}
where $E_\text{opt} = |U| (2 S - 1) + |V_b|$ corresponds to the energy difference between the ground state and the first excited state of $\langle 0| H_0 | 0\rangle$ at $\Delta \to 0$, $|0\rangle$ is the photon vacuum.
In order to see optical transitions between the ground state and the second excited state of $\langle 0| H_0 | 0\rangle$, two virtual pseudospin flips are required, such transitions emerge in order $\propto \Delta^4$.
We indeed observe such transitions in exact diagonalization, they are strongly suppressed compared to the leading harmonic, see Fig.~\ref{fig:optical_cond_2nd_order_QPT} in the main text.
In order to find the Fourier transform $\Pi_0(g, \omega)$, we use the Bessel function expansion,
\begin{eqnarray}
&& e^{z \cos(\omega_0 t)} = \sum\limits_{-\infty}^\infty I_m (z) e^{i m \omega_0 t} , \label{Bessel}
\end{eqnarray}
where $I_{-m}(z) = I_{m} (z)$ is the modified Bessel function of the first kind.
The real part of the optical conductivity then follows from Eq.~(\ref{Pi0factor}),
\begin{align}
\label{sigT}
&{\rm Re}\left[\sigma(\omega)\right] \approx -\frac{1}{\omega} {\rm Im}\left[\Pi_0(g, \omega)\right]=\\
&\frac{\pi S}{2 \omega} \left(e b \Delta\right)^2 \sum\limits_{n = 0}^\infty p_n\left(g^2\right) \nonumber\\
&\cdot\sum\limits_{m \in \mathbb{Z}} e^{-2 g^2 N_\text{ph}} I_m \left(2 g^2 N_\text{ph}\right) \delta \left(\omega - E_\text{opt} - (n - m) \omega_0 \right) \nonumber, 
\end{align}
where $\mathbb{Z}$ is the set of integers and $p_n(z) = e^{-z} z^n/n!$  the Poisson distribution.
Notice that at $T = 0$ we get $N_\text{ph} = 0$, so only the $m = 0$ term in Eq.~(\ref{sigT}) contributes, and we restore Eq.~(\ref{siganalytic}) in the main text. The subleading $\propto \Delta^4$ harmonics can be calculated similarly via perturbative expansion with respect to the terms $\propto \Delta$ in $H$, see Eq.~(\ref{eq:Peierls_Hamilt}) in the main text.
Here we only present the brightest harmonics $\propto \Delta^2$.

\begin{figure}[h]
		\includegraphics[width=0.76\columnwidth]{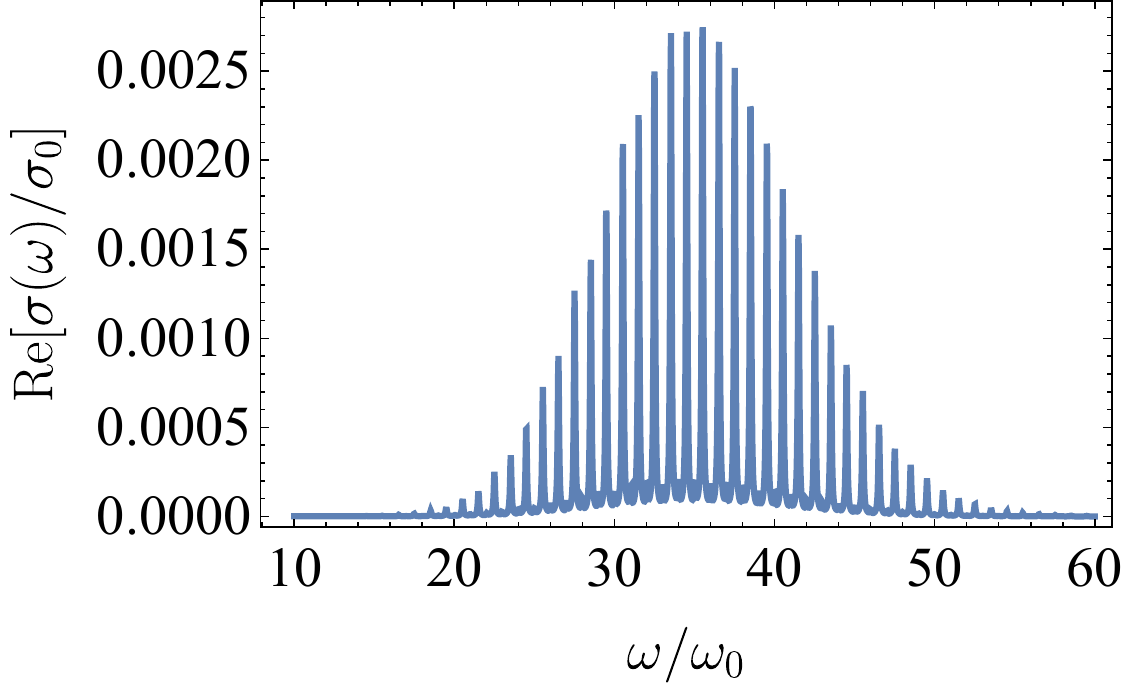}\put(-186,109){(a)}\\
		\includegraphics[width=0.76\columnwidth]{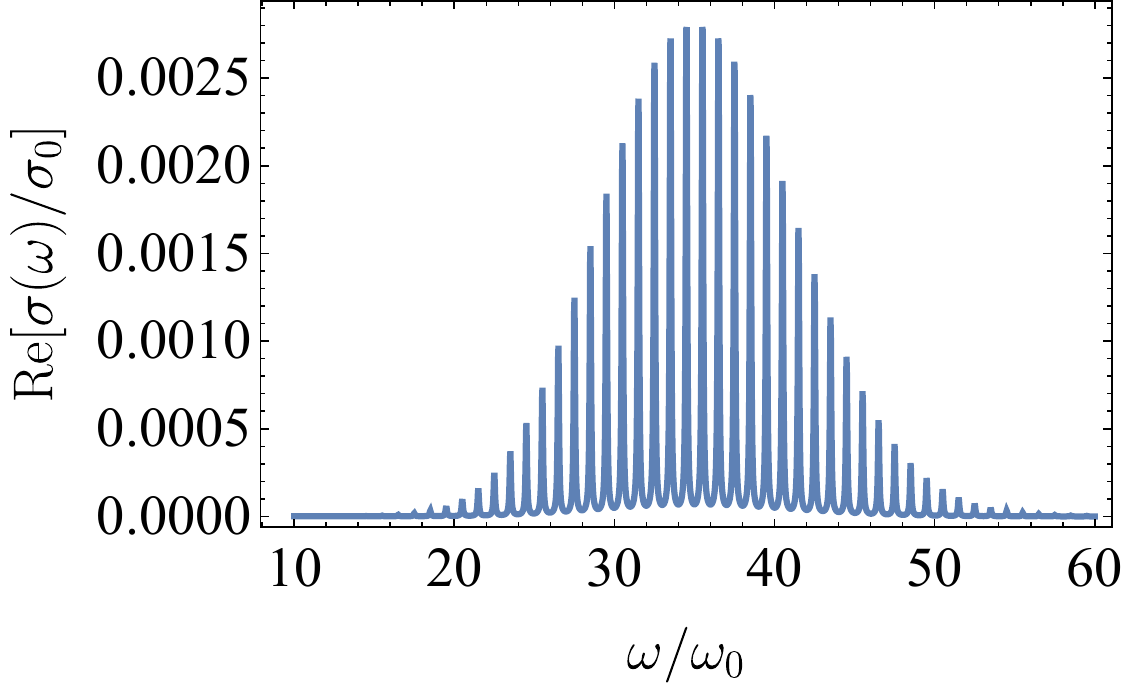}\put(-186,109){(b)}
	\caption{\label{fig:opt_cond_num_vs_analytics} Comparison between the numerics (exact diagonalization, (a)) and analytics (Eq. (12) from the main text, (b)) for $N = 2$ DQDs at $U/\omega_0=-0.5$ and $\omega_0/\Delta=1$. }
	\vspace{-5pt}
\end{figure}

Interestingly, the Poissonian structure of the frequency comb in $\operatorname{Re}\left[\sigma(\omega)\right]$ is similar to the down-conversion in circuit QED~\cite{PhysRevLett.130.023601, Mehta2023}, and to the replica bands recently discussed in the context of light-matter interaction~\cite{Eckhardt2022}.
In Fig.~\ref{fig:opt_cond_num_vs_analytics} we show the comparison between the optical conductivity $\operatorname{Re}[\sigma(\omega)]$ calculated via exact numerical diagonalization and the analytical result (Eq. (\ref{siganalytic}) from the main text) in the ordered phase.

\bibliography{bibliography}

\end{document}